# Active Control of Polariton-Enabled Long-Range Energy Transfer

A. Cargioli[1,2,*,#], M. Lednev[3,#], L. Lavista[1,2,#], A. Camposeo[2], A. Sassella[4], D. Pisignano[2,5], A. Tredicucci[2,5], F. J. Garcia-Vidal[3], J. Feist[3], and L. Persano[2,†]

[1]Dipartimento di Fisica "E. Fermi", Università di Pisa, Largo B. Pontecorvo 3, I-56127 Pisa, Italy

[2]NEST, Istituto Nanoscienze-CNR and Scuola Normale Superiore, I-56127 Pisa, Italy

[3]Departamento de Física Teórica de la Materia Condensada and Condensed Matter Physics Center (IFIMAC), Universidad Autónoma de Madrid, E-28049 Madrid, Spain

[4]Dipartimento di Scienza dei Materiali, Università degli Studi di Milano-Bicocca, Via Roberto Cozzi 55, I-20125 Milano, Italy

[5]Dipartimento di Fisica "E. Fermi" and Center for Instrument Sharing (CISUP), Università di Pisa, Largo B. Pontecorvo 3, I-56127 Pisa, Italy

[*]Contact Email (present address): acargioli@phys.ethz.ch

[†]Contact Email: luana.persano@nano.cnr.it

[#] These authors contributed equally to this work

**Abstract**

Optical control is achieved on the excited state energy transfer between spatially separated donor and acceptor molecules, both coupled to the same optical mode of a cavity. The energy transfer occurs through the formed hybrid polaritons and can be switched on and off by means of ultraviolet and visible light. The control mechanism relies on a photochromic component used as donor, whose absorption and emission properties can be varied reversibly through light irradiation, whereas in-cavity hybridization with acceptors through polariton states enables a 6-fold enhancement of acceptor/donor contribution to the emission intensity with respect to a reference multilayer. These results pave the way for synthesizing effective gating systems for the transport of energy by light, relevant for light-harvesting and light-emitting devices, and for photovoltaic cells.

In the strong light-matter coupling regime, photons confined within an optical cavity interact with material emitters, thus changing the fundamental physical properties of the coupled system and creating hybrid light-matter states[1,2]. Excitations of these states are quasiparticles named polaritons, carrying features of both photons and excitons. One consequence of polariton formation is that the energy spectrum of the system changes, featuring two peaks separated, at zero cavity-transition detuning, by the Rabi splitting. The potential to modify material properties and chemistry underneath through strong light-matter coupling has stimulated enormous interest from the scientific community, both at the fundamental level and for its potential technological applications[3-5]. Organic materials provide relevant opportunities in this context, due to their large oscillator strengths that can lead to the achievement of large Rabi splitting values. Frenkel excitons[6] might strongly localize in organics at single-molecule level, with binding energies of the order of 1 eV[7], and Rabi splittings of hundreds of meV might enable the observation of macroscopic quantum phenomena at room temperature. In this framework, some remarkable achievements include room temperature Bose-Einstein condensation[8], polariton lasing[9,10], tunable third harmonic generation[11], and increased efficiency in organic photovoltaics (OPVs)[12]. Indeed, one of the main challenges in OPVs is the improvement of the power conversion efficiencies (PCE)[13,14], that suffer from the relatively large non-radiative decay rates and the typically incoherent, diffusive nature of exciton transport. The formation of delocalized polaritons in the collective strong coupling regime, which originates from the photonic component, has the potential to enhance energy transfer efficiencies overcoming low exciton transport and charge carrier mobility, thus effectively leading to an improvement of the overall efficiency of light harvesting[12]. The long-range energy transfer offered by polariton states already led to a promising outlook for the enhancement of the PCE[15].

In conventional Förster-type energy transfer processes, energy transport is based on exciton dipole-dipole interactions between a donor and an acceptor molecule, with a low effective range of a few nm[7]. This usually requires physical blending of different molecular components to enable energy transport. Instead, in the strong coupling regime the quantum-mechanical entanglement of the donor and acceptor molecules within the polaritonic states enables a new energy transport mechanism that is no longer dependent on the spatial distance[16,17]. Several reports indicate that mixed exciton-polariton states serve as fast pathways for the energy transfer from the donor molecules to the acceptor ones[18-20] and, consequently the spatial range of transfer has been extended from 10 nm[21] to a few micrometers[16,22]. These results have been obtained by physically separating the donor and acceptor molecules by embedding them in layered systems with a transparent spacer[23]. In general, long-range transfer is not only limited to molecular systems, but has been also demonstrated with carbon nanotubes excitons[24-25] and vibrational excitations[26]. However, such systems are basically static. While the coupling parameters are traditionally permanently defined by a given cavity design (i.e. layer composition, thickness, and topology), dynamic systems where external stimuli might activate or deactivate polariton states[27-30] would open much more exciting perspectives for precisely controlling energy flows in intelligent resonant photonics.

Here we propose a new class of optical cavities based on a photogateable donor-acceptor system, in which UV-light driven photoisomerization directly affects the energy transfer mechanism. A microcavity architecture is developed by two different dye layers, sequentially deposited from orthogonal solvents to form the donor-acceptor system. UV light irradiation activates the photoisomerization process of the donor, thus controlling the concentration of transfer-available components in the cavity. As the concentration increases, polariton states are formed and the energy transfer process to the acceptor is activated. Furthermore, irradiation with visible light switches back the energy levels to the initial uncoupled conditions, thus deactivating the polariton-assisted energy transfer process. The capability to control complex energy flows in photonic devices by means of external light provides additional functionalities and opportunities in light harvesting based on strong light-matter coupling.

## Results and Discussion

The microcavity architecture, schematised in Fig. 1a, consists of two Ag films as mirrors, sandwiching two spatially separated photoactive layers.

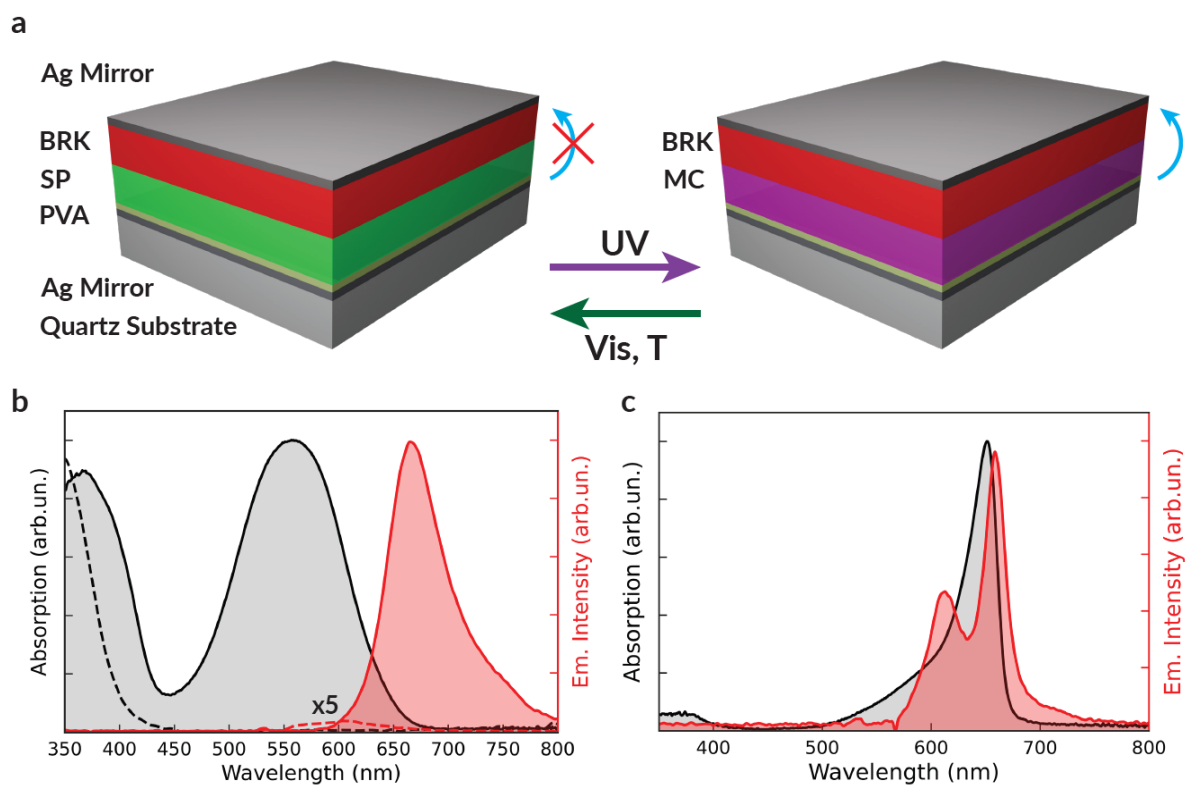

**Fig. 1: Device architecture and molecular system. a** Schematics of the cavity before and after photochromic donor conversion. The donor molecule, initially transparent in the visible range in its SP form, is converted to a colored MC form by irradiation with UV light (violet arrow), whereas the back-conversion can occur by irradiation with green light (green arrow) or by thermal relaxation. The vertical bent arrows represent the donor-acceptor energy transfer in the two configurations. **b** Absorption spectrum of a PMMA film with SP (black dashed lines) and MC (black continuous line) and corresponding PL spectrum of the SP (×5 intensity, red dashed line) and MC form (converted by UV exposure for 5 s, red continuous line). **c** Absorption (black line) and emission (red line) of a film of PVA doped with BRK. The excitation wavelength for the emission measurements is 532 nm.

The absorption and emission of the donor layer, which is based on the photochromic 1,3,3-Trimethylindolino-6'-nitrobenzopyrylospiran (SP)[28] in a host matrix of poly(methyl methacrylate) (PMMA), are reversibly varied by UV and green light irradiation. The SP film is transparent in the 450-800 nm range (Fig. 1b) and exhibits a highly uniform morphology (root mean square roughness = 0.3 nm, Fig. S1a,b). Upon irradiation with UV light ($\lambda_{UV}$ = 365 nm), SP converts to merocyanine (MC). For each value of the duration of the UV exposure, a mixture of SP and MC is obtained, with relative content depending on the specific irradiation conditions and only the MC component being coupled to the cavity mode. MC features a strong absorption peaked at 554 nm, while its photoluminescence (PL), measured with a pump laser at 532 nm, is peaked at 663 nm (Fig. 1b). The PL from the PMMA-SP film under the same excitation conditions features only a very weak peak at 600 nm (Fig. 1b), in agreement with previous reports[31].

The controlled photochromic conversion is exploited to activate/deactivate the coupling to the microcavity, and the resulting excited state energy transfer to the acceptor molecules. The length of the active region is chosen to have the second-order resonant mode at about 620 nm at normal incidence (inter-mirror distance = 355 nm), a configuration that is expected to enhance the light-matter coupling for both the donor and acceptor molecules[16,23] (Fig. S2) , and additionally minimizes interface effects between the two layers as the field has a local minimum there. The J-aggregate[32] form of 3,3'-Bis(3-sulfopropyl)-4,5:4',5'-dibenzo-9-ethylthiacarbocyanine betaine thiethylammonium salt (BRK)[16,19] is used as acceptor, embedded in a host matrix of polyvinyl alcohol (PVA). BRK absorbs at 655 nm, whereas its emission shows peaks at 612 nm and 659 nm, respectively (Fig. 1c). The most intense peak is attributed to fully formed J-aggregates, while the smaller and blue-shifted one is traceable to some BRK molecules which do not aggregate in the fabrication process (see Supplementary Information -SI-, Fig. S3). While contributing to BRK absorption broadening, non-aggregated molecules do not interfere severely with the polariton formation since their number, and thus their coupling to the cavity field, is sufficiently small compared to the fully formed aggregates. The kinetics of the SP-to-MC photochromic conversion upon exposure to 365 nm light is shown in Fig. S4. Further details on microcavity fabrication are reported in the Methods, Section 2 of SI and Fig. S5.

The optical transmission properties of the microcavity before and after 180 s of UV exposure are illustrated in Fig. 2. Optical transmission measurements at intermediate UV exposure times are reported in Fig. S6. The experimental data are compared with simulated transmission maps, computed through a transfer matrix approach[33]. To this aim, the wavelength dispersion of the complex refractive index of PMMA-MC and PVA-BRK are derived from the optical transmission measurements performed on reference first-order cavities embedding either BRK-doped PVA or MC-doped PMMA, respectively (details in Section 4 of SI, Fig. S7 and S8). In pristine devices (UV exposure time = 0 s), only the BRK aggregates couple to the cavity field at visible wavelengths. Since the BRK absorption is off resonance to the cavity dispersion, we observe only a slight shift with respect to the uncoupled bands (the resulting light-matter coupling constant for BRK is $g_{BRK} = 117$ meV). Once MC is introduced in the system by means of UV irradiation, two exciton species can participate in the polariton formation and three polaritonic branches appear (Fig. 2), i.e., the upper polariton branch (UPB) at about 503 nm, the middle polariton branch (MPB) at 612 nm and the lower polariton branch (LPB) at 675 nm (all wavelengths at normal incidence). A fit of the polaritonic dispersions is also performed using a coupled oscillator model[34], which allows the Hopfield coefficients to be retrieved for each polariton branch (see Fig. S9, S10, and S11). The results of this analysis at 0° are shown in Fig. 3. As expected[16,19,34], the UPB (LPB) is mostly composed of a photonic component and a donor (acceptor) molecular component, while the MPB has a more balanced nature involving the three components. By increasing the exposure time to UV light, the Hopfield coefficient of MC is increased up to about 0.2 for the MPB, while its cavity component is lowered down to about 0.4. The MPB evidences the possibility to control the degree of hybridization between the donor and acceptor molecules by an external light signal. We also mention that in between the polaritonic branches visible in the transmission spectra, the system also contains "dark states" or "exciton reservoirs" corresponding to the excitonic transitions of the donor and acceptor molecules that are not coupled to the cavity field.

The light-matter coupling strength is known to depend on the square root of number of emitters interacting with the cavity field[35,36]. Since we are actively changing the concentration of donors available to energy transfer (MC), we expect the light-matter coupling constant to increase with UV-exposure time ($t_{exp}$). Through the coupled oscillators model (see Methods and SI for details), we extract the parameter $g_{MC}(t_{exp})$ from the transmission measurements. As shown in Fig. S12, an increase of $g_{MC}$ up to 124 meV is found after 180 s of UV irradiation. The system can be back-switched[37] by irradiation with green laser light (532 nm, intensity $\sim$ 275 mW cm$^{-2}$), which reverses the SP-MC photoisomerization (Fig. 4).

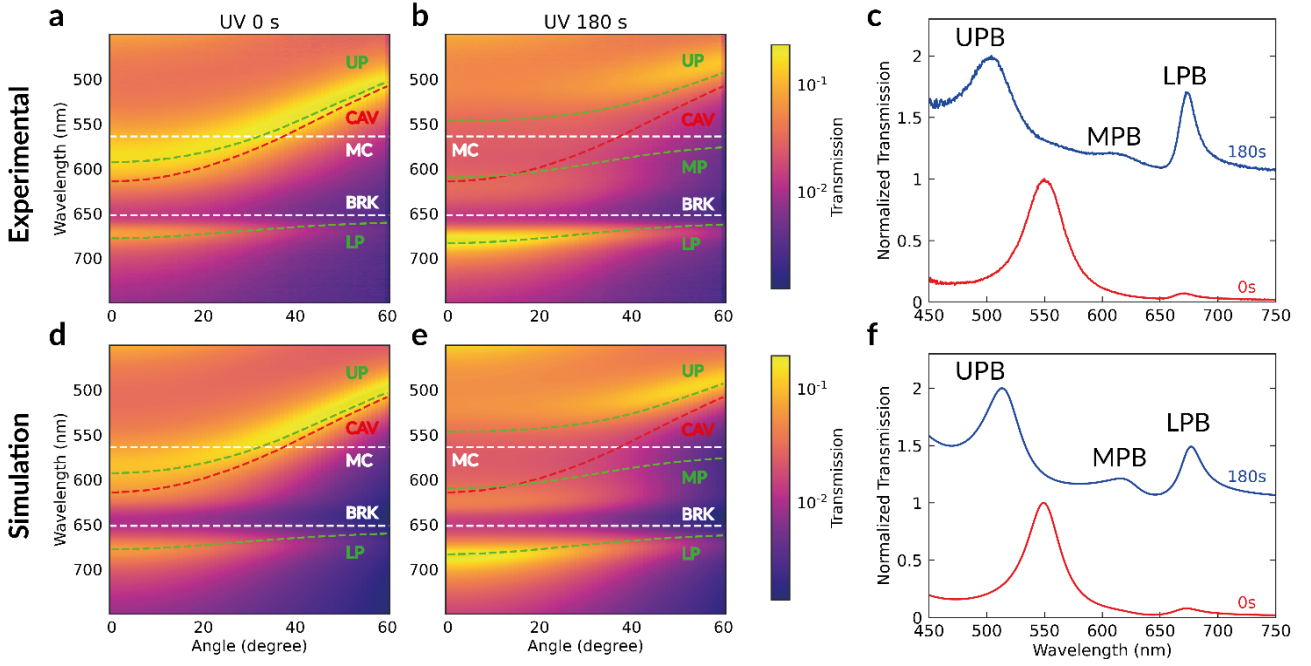

**Fig. 2: Angle-resolved transmission. a,b:** Experimental angle-resolved transmission maps at different UV exposure times, and transmission spectra (**c**) before (red line) and after (blue line) irradiation at the anticrossing angle (37°). **d,e,f:** corresponding simulated maps and transmission spectra. In the colorscale, unity stands for total transmission. In each colormap the bare cavity mode (red dashed line), the MC excitonic transition (upper white dashed line) and the BRK excitonic transition (lower white dashed line) are also reported. The green dashed lines are the result of a fit using the coupled oscillators model[30]. The transmission spectra shown in **c** and **f** are divided by their maximum value, respectively, and vertically shifted for better clarity.

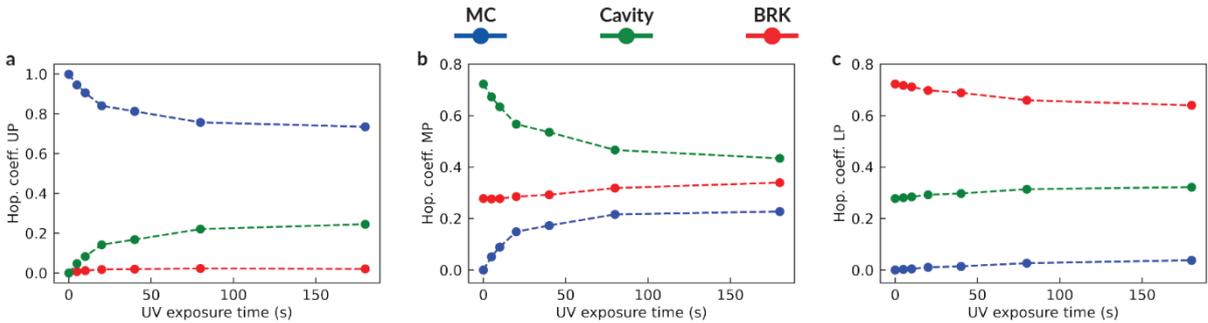

**Fig. 3: Hopfield coefficients.** Hopfield coefficients at 0° of upper **a**, middle **b**, and lower **c** polariton branches as a function of the UV exposure time.

The transmission maps for various times of green light exposure are reported in Fig. S13 and S14. Ultimately obtained bands are largely comparable to those of the pristine device, with minor changes of signal intensity and broadening of the photonic mode most likely due to the residual MC component. The complete set of polariton branches can be observed in the system for up to four consecutive UV-green irradiation cycles (Fig. S15). Fatigue effects, attributed to photo-oxidation[38], photoisomerization towards poorly back-converting forms[39], or formation of MC aggregates[40], then lead to MPB suppression. Lack of photoconversion is found for ten or more irradiation cycles in the here investigated systems. Various strategies have been developed to reduce fatigue in SP/MC compounds, including covalent attachment of modified SP to polymer films[41] and embedment of sulfonated SP in silica matrix[42].

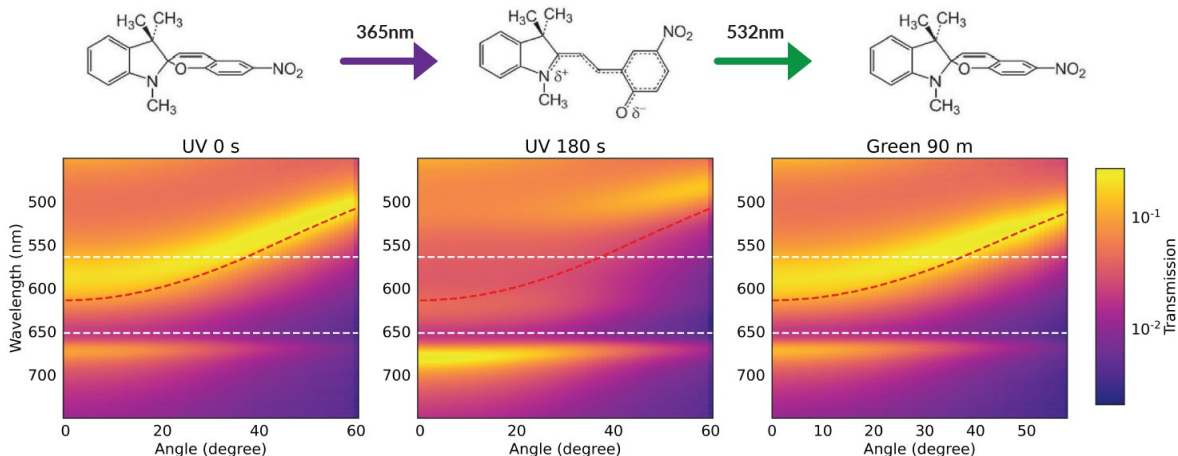

**Fig. 4: Polariton switching.** Angle-resolved transmission measurements as a function of UV and green light exposure time. The photochromic conversion from SP to MC and back to SP is also schematically displayed. In each colormap, the bare cavity mode (red dashed line), the MC excitonic transition (upper white dashed line) and the BRK excitonic transition (lower white dashed line) are also reported.

The angle-resolved emission from cavities after different UV exposure times and then shortly excited by a 532 nm laser are shown in Fig. 5. The corresponding emission spectra are reported in Fig. S16. The emission spectrum of the cavity before irradiation with UV light shows two bands peaked at 600 nm and 668-671 nm, respectively. As soon as the UV light is switched on, the upper band intensity decreases and finally disappears while the lower one red-shifts and its intensity increases. In agreement with the transmission data, a back-conversion of the PL signal is found upon longer (5-90 min) irradiation with green light (Fig. S17 and S18).

The observed emission spectrum can be rationalized from a model in which the emission from the molecular reservoirs to the outside is calculated using transfer matrix theory, which can be conceptually understood as the polaritonic modes behaving as a filter for the emission from the molecular reservoir. Although the emitted light is transmitted from within the cavity to the outside, the relevant filter function can be well-approximated as the conventional cavity transmission function (see Section 9 and Fig. S19 in SI). This does not imply that the cavity and polariton formation has no effect apart from acting as a cavity filter, since the cavity and polariton modes furthermore act to mediate efficient donor-acceptor energy transfer as discussed in the following. Thus, we represent the emission as the PL signal of the bare molecules modulated by the cavity transmission:

$$I_{cav}(\omega, \theta, t_{exp}) = T_{cav}(\omega, \theta, t_{exp})[\alpha(t_{exp})I_{BRK}(\omega) + \beta(t_{exp})I_{MC}(\omega, t_{exp})] \tag{1}$$

where $T_{cav}$ is the cavity transmission of the hybrid system, $I_{BRK}$, $I_{MC}$ are the emission intensities of the molecules outside the cavity, and $\alpha$, $\beta$ are phenomenological weight coefficients. The coefficients effectively represent the contributions of both molecular species to the emission of the hybrid system. For the simulation of the emission maps, we fit $I_{cav}$ (Eq. 1) to the experimental PL intensity from the cavity, using the weight coefficients $\alpha$, $\beta$ as free parameters. This approach for the simulation of the emission properties of the cavity is equivalent to a rate equation model, which has been successfully applied for the interpretation of the emission measurements in similar systems[34,43,44] under the assumption that radiative pumping[45] is the dominant population mechanism for the LPB, i.e., the vibrational scattering from the acceptor excitonic reservoir is negligible (see Section 10 and Fig.s S20-S21 of SI for details). In our hybrid system, the used approximation is valid since radiative pumping occurs not only from the acceptor reservoir, but also from the donor one, due to the fact that the MC emission has a significant overlap with the LPB. The results of the simulations are shown in the bottom of Fig. 5. As an example of validity of the model, we report in Fig. 6a the comparison between the measured and simulated integrated emission intensity of the LP at a fixed angle (in this case 0°) as a function of the UV exposure time.

The simulated data reproduces the experimental measurements well. They demonstrate that the increase in the number of MC molecules not only dramatically changes the absorption properties of the system but also its emission, with a direct impact on the energy transport from donor to acceptor molecule. Thus, we can fully interpret the emission spectra dynamics of the hybrid system under UV illumination. At 0 s of UV irradiation, the upper band is attributed to the emission of non-aggregated BRK molecules which leaks through the cavity mode slightly modified by interaction with BRK molecules. After the start of the UV illumination the hybrid states shift due to an increase in the donor coupling strength. This, in turn, leads to a quenching of the upper emission band, since the emission peak of the BRK molecules is not in resonance with the transmission bands of the hybrid system. For the lower emission band the situation is completely different, since for the whole range of UV irradiation times of the measurements the LPB remains in resonance with the main emission peaks of the donor and acceptor molecules. Once the donor is introduced with varied amounts in the system by UV irradiation, the emission of the multilayer outside and inside the cavity become remarkably different (Fig. S22, S23): outside the cavity both donor and acceptor molecules contribute to a broad and spectrally-stable overall emission, whereas for the cavity a red-shift (~8 nm) of the emission peak is found upon increasing the UV exposure time. It is worth noting that also control cavities involving either only the acceptor (Fig.s S24-S26) or only the donor (Fig.s S27-S29) show a substantially different behavior. The transmission and emission of the acceptor-only cavity are almost unaffected by the UV irradiation. Instead, in the donor-only cavity polariton bands still vary upon UV irradiation due to the photo-induced SP-to-MC conversion, with emission occurring at a different wavelength range with respect to the donor-acceptor cavity (i.e., at 648-658 nm, related to the specific LP formed in presence of the unique PMMA-SP/MC layer).

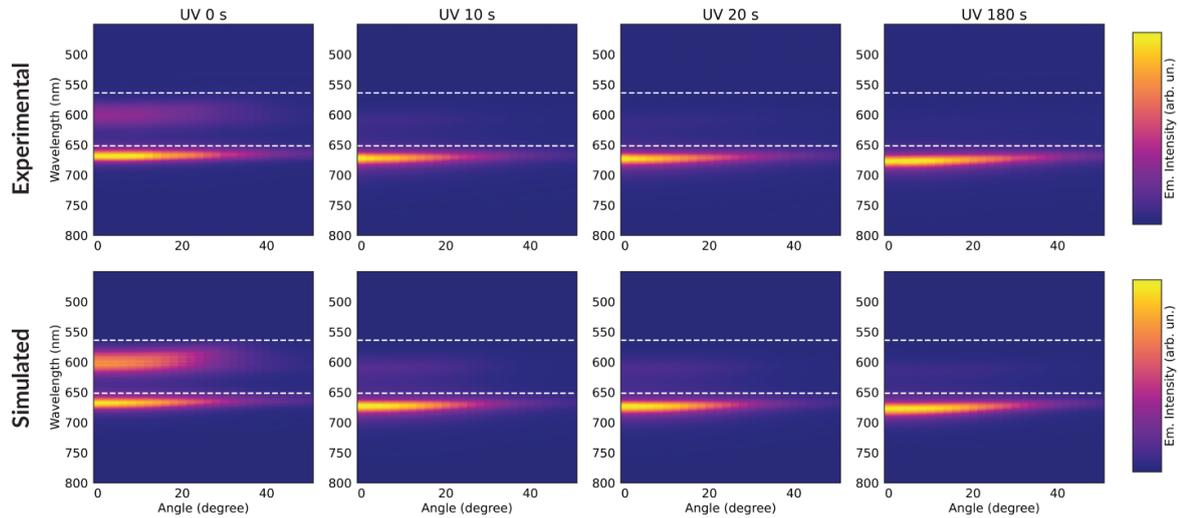

**Fig. 5: Angle-resolved PL.** Experimental and simulated angle-resolved emission maps of the cavity after UV exposure time of 0 s, 10 s, 20 s and 180 s (from left to right) and then excited by a 532 nm pump (< 1 min). White dashed lines show the spectral wavelengths of donor and acceptor absorption peaks.

Our approach allows us not only to clearly understand the origin of the observed experimental PL features, but also to track the contributions of both molecular species to the emission through the weight coefficients in Eq. 1. Following an analogous procedure, we find the weight coefficients corresponding to BRK and MC molecules for emission of the multilayer outside the cavity (for details see SI section 14, Fig. S30). The comparison between the normalised weight coefficients for the donor and acceptor molecules placed outside and inside the cavity is reported in Fig. 6b-c, respectively (emission spectra for the PMMA-MC/PVA-BRK multilayer without cavity are shown in Fig. S23a). The normalised weight coefficients are defined as $\alpha_n = \frac{\alpha}{\alpha+\beta}$ and $\beta_n = \frac{\beta}{\alpha+\beta}$, corresponding to the relative fractions of emission arising from the BRK and MC molecules, respectively, which cannot be derived by simply comparing the integrated emission intensities of the cavity and reference multilayer (Fig. S23b). Moreover, the normalisation of the coefficients is necessary to properly compare the emission properties of the multilayer outside and inside the cavity, and to rule out any dependence on the excitation efficiency which can significantly change throughout the experiments in the cavity due to the shift of the polaritonic states (see Sections 14,15 of SI). While in the transmission measurements, the polariton bands reach their final positions only when the reaction reaches a steady state, the emission becomes stable much more quickly, after just a few seconds of irradiation. At $t = 0$, emission is mainly from BRK. In the first few seconds of UV irradiation and donor photoisomerization, the relative amount of the BRK-related emission fraction, $\alpha_n$, is strongly reduced as a consequence of the sudden rise of MC concentration. Without the cavity (Fig. 6b) this effect is very pronounced and slowly continues afterwards, consistent with the progressive achievement of the photostationary state, until $\beta_n \sim 1.5\ \alpha_n$.

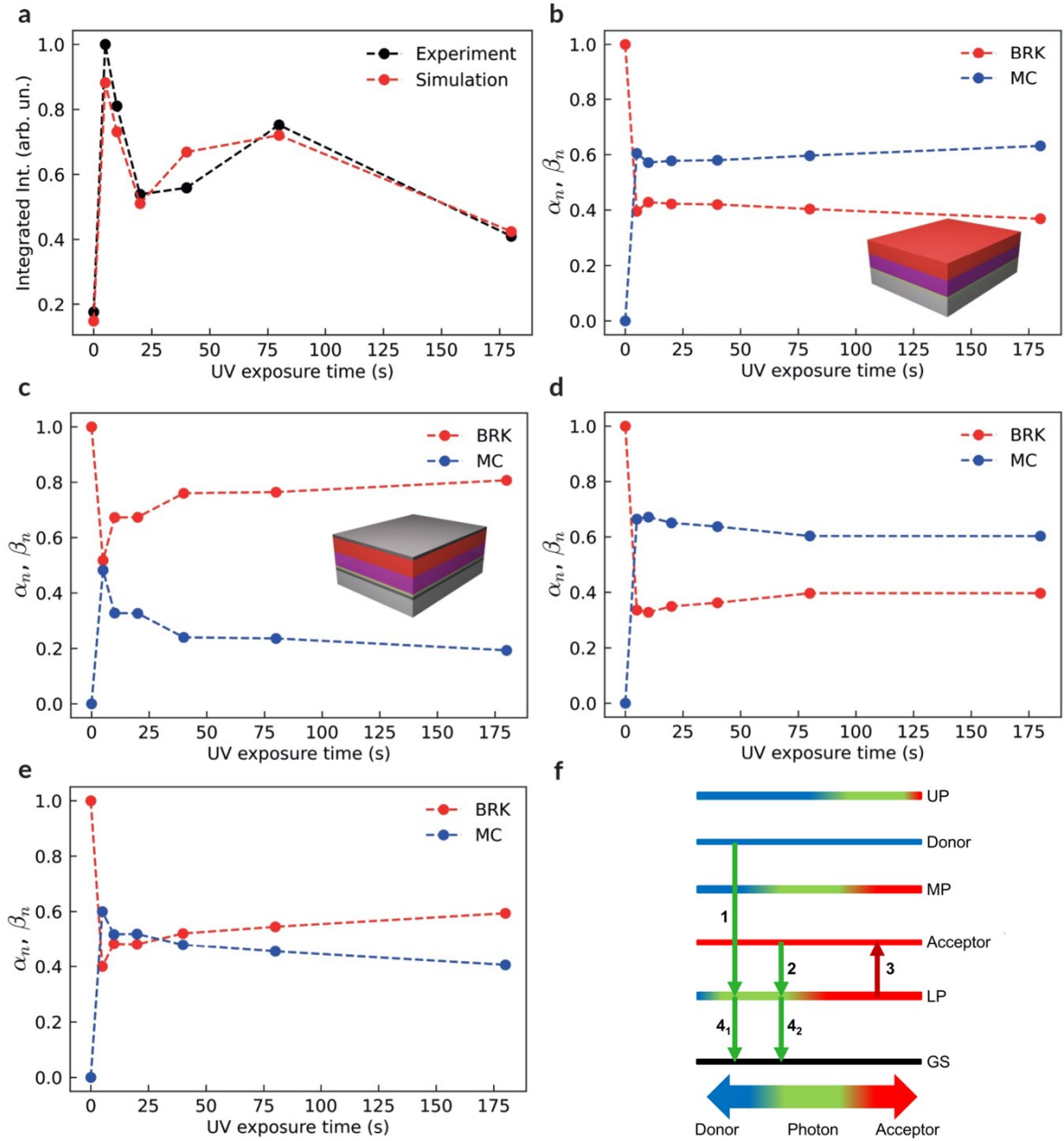

**Fig. 6: Modelling the light-controlled energy transfer mechanism. a** Simulated (red circles) and measured (black circles) integrated emission intensity of the LP at 0° as a function of the UV exposure time. Experimental and simulated data are divided by the maximum value of the experimental data. **b-c** Normalised weight coefficients, $\alpha_n$ (red circles) and $\beta_n$ (blue circles), as a function of UV exposure time for the MC/BRK system outside and inside the cavity, respectively. **d-e** $\alpha_n$ (red circles) and $\beta_n$ (blue circles) for the off-resonance MC/BRK system outside and inside the cavity, respectively. **f** Schematic representation of the relevant energy levels and of the emission decay pathway for the cavity analyzed in **a-c**.

On the other hand, inside the cavity, the roles are reversed and at $t_{exp} > 5$ s the BRK contribution rises and achieves a value almost 4 times higher than the MC, due to the presence of cavity-enhanced energy transfer. We additionally realize an off-resonant cavity, where the second-order resonant mode is red shifted (772 nm) by increasing the thicknesses of the donor and acceptor layers (see Methods for details). By doing so, the cavity mode is out of resonance from the excitonic states of the molecular species. The resulting measured angle-resolved transmission and PL spectra are reported in Section 16 of the SI (Fig.s S31-S34), while the calculated normalised weight coefficients for a donor/acceptor multilayer out of the cavity and inside the cavity are shown in the Figure 6d,e, respectively. For the reference multilayers out of cavity, the contribution to emission from MC and BRK are similar for resonant and non-resonant configurations (Fig. 6b,d). Interestingly, in the off-resonance cavity, the BRK and MC almost equally contribute to the overall emission up to about 40 s of UV exposure, whereas at $t_{exp} > 40$ s the BRK contribution slightly increases reaching a value which is significantly lower than in the resonant case ($\alpha_n \sim 1.5\ \beta_n$).

We explain this behavior through the level scheme in Fig. 6f. Firstly, the green laser pumps both the donor and acceptor excitonic reservoirs, then the molecules emit into the lower polariton state through its photonic component (radiative pumping mechanism corresponding to arrows 1 and 2 in the scheme). The lower polaritonic state has two main loss mechanisms: the dominant one is radiative decay through the cavity mirrors (arrows $4_1$ and $4_2$ for light originating from the donor and acceptor reservoirs, respectively, which can be distinguished with the fitting procedure discussed above, and which are characterized by the coefficients $\beta$ and $\alpha$, respectively), which occurs on femtosecond timescales, while the second one is non-radiative decay to the acceptor excitonic reservoir (arrow 3), with efficiency proportional to acceptor fraction in LP. While this process is usually expected to be slower than radiative decay[34, 43, 44, 46], transfer matrix calculations of the relative absorption and emission fractions for light emitted by the donor molecules indicate that this process is actually comparably efficient to radiative decay in the current setup (see Section 10, Fig. S20), which can be attributed to the relatively large overlap between the acceptor absorption spectrum and the lower polariton[46]. Its appearance significantly affects the excitation transfer pathways due to pumping of the acceptor excitonic reservoir. In particular, it provides transfer of energy from the donor reservoir to the acceptor one through the lower polaritonic state. Thus, the lower polaritonic state in our experimental setup serves as an intermediate state for energy transfer between the donor and acceptor, and is responsible for the redistribution of the donor and acceptor contributions to the emission (Fig. 6b-c). Overall, in the resonant cavity the fraction of the emission due to the acceptor molecules with respect to the donor ones is enhanced by a factor of 6 compared to the bare donor/acceptor multilayer. By contrast, in the off-resonant cavity the redistribution between donor and acceptor is weakened because of the reduced efficiency of (i) the radiative pumping from the donor excitonic reservoir to the LP state (arrow 1 in Fig. 6f) due to a decrease of the overlap between the LP dispersion and MC emission band and, (ii) the non-radiative relaxation of lower polaritons to the acceptor excitonic reservoir (arrow 3 in Fig. 6f) since the efficiency of this channel is proportional to the acceptor fraction in the LP, which is largely decreased. Indeed, for the off-resonant case, the LP consists mostly of the photonic part (see the calculated Hopfield coefficient of the LP in Fig. S35).

In conclusion, we have demonstrated the possibility of controlling the polariton formation between two different molecules via external optical gating in a donor-acceptor system. This is achieved by embedding a photo-active multilayer in an optical microcavity, in which one of the layers (the donor one) features reversible photochromic properties upon UV and visible light irradiation. These architectures enable the possibility to control the energy transport between the spatially separated species by light. Engineering externally controllable, intelligent photonic systems which could present long range energy transport might open a new way of approaching light-harvesting, light-emitting, and OPV devices and integrated platforms.

**Methods**

**Microcavity fabrication**. PMMA and SP with a 1:1 weight:weight (w:w) ratio are dissolved in toluene, while PVA and BRK (10:1 w:w ratio) are dissolved in a mixture of deionized water and methanol (1:1 volume ratio). Microcavities are fabricated by thermal evaporation of a 25 nm-thick Ag mirror on top of a quartz substrate ($1 \times 1$ cm$^2$) by using a MBRAUN MB-ProVap 4G system. Afterwards, the active organic multilayer is deposited onto the Ag mirror by spin coating a PVA buffer layer (thickness: 25 nm) as a first step, followed by a PMMA-SP layer (thickness: 150 nm) and a PVA-BRK layer (thickness: 180 nm). Finally, a 25-nm thick top Ag mirror is evaporated onto the PVA-BRK layer (SI, Fig. S5). The thicknesses of the active organic layers in the control cavities are: donor-only cavity, PVA (25 nm), PMMA-SP (150 nm), PVA (180 nm); acceptor-only microcavity, PVA (25 nm), PMMA (150 nm), BRK-PVA (180 nm); off-resonant cavity, PVA (25 nm), PMMA-SP (225 nm), PVA-BRK (240 nm). The reference PVA/PMMA-SP/PVA-BRK multilayers are deposited by spin coating on top of a quartz substrate. The thicknesses of the Ag and organic layers are measured by using a stylus profilometer (DektakXT, Bruker). The surface morphology of PMMA-SP films is investigated in PeakForce tapping mode by using a probe with a nominal spring constant of 0.4 N m$^{-1}$ (Bruker, USA) on a Bruker Dimension Icon system, equipped with a Nanoscope V controller. More details about the fabrication are reported in the SI.

**Optical characterization and light switching**. The absorption and transmission spectra of the Ag, PVA-BRK and PMMA-SP/MC layers are measured by using a spectrophotometer (Lambda950, Perkin Elmer). Spectroscopic ellipsometry measurements are performed on PVA-BRK films spin coated on a Silicon/silicon oxide substrate using the V-VASE ellipsometer (J. A. Woollam Co.) in the spectral range from 300 to 800 nm and with three angles of incidence, 70°, 75°, and 80°. The fitting procedure is carried out by means of the software WVASE32, considering a multiple-oscillator model. PL spectra of the active layers are measured by exciting the samples with a 532 nm diode-pumped solid-state laser and analyzing the emission by using a fiber-coupled monochromator (FLAME, Ocean Optics). Photochromic conversion experiments are carried out by irradiating the whole surface of the samples with a UV light emitting diode (LED, mod. M365LP1, Thorlabs, emission peaked at about 365 nm) for the conversion of SP in MC, while the back-switching is achieved by illuminating the samples with the 532 nm laser.

**Angular transmission and PL measurements**. Angle-resolved transmission measurements are performed by using the output beam of a broadband lamp (DH-2000, Ocean Optics) focused onto the sample (diameter of the spot about 0.5 mm at normal incidence). The spectrum of the lamp in air is taken as reference. A polarizer is positioned along the optical path to control the polarization of the incident light beam. The sample is placed on a holder mounted on one of two concentric rotation stages, used for varying the angle of the sample with respect to the incident beam and the angle of collection of the optics, respectively. For angular transmission measurements, the collection optics (composed by a lens system and an optical fiber) is positioned along the axis of the incident beam, while the sample is rotated. The light collected by the lens system and the optical fiber is directed to the monochromator for spectral analysis. For angular PL measurements the sample position is fixed, while the collection optics is rotated. The samples (both cavities and reference films) are excited by the 532 nm laser, impinging on the sample with an incidence angle of about 5° (spot size about 0.5 mm). The PL of the samples is collected by the lens system and the optical fiber positioned on the rotation stage, and a long-pass filter (cut-off wavelength at 550 nm) is used to attenuate the light of the excitation laser. In a typical measurement, PL angular spectra are collected with a step of 2°, while the angle of collection of the PL is about 2.5°. The excitation intensity and the exposure time of the green light during PL measurements were reduced to 19 mW/cm$^2$ and 30s, respectively, not to affect the MC to SP back-switching.

**Modelling**. For the simulation of the angle-resolved transmission spectra, TMM is used[47]. Firstly, in order to extract the refractive indices of the active layers (PVA-BRK and PMMA-SP(MC)) we carry out TMM calculations for first-order cavities containing only a PVA-BRK layer and only a PMMA-SP(MC) layer, respectively (details in SI, Section 4). Then, the full multilayer cavity with both donor and acceptor layers is considered. Using TMM, we fit the simulated spectra to the experimental ones for exposure times $t_{exp}$= 0, 5, 10, 20, 40, 80, 180 s, where the fit parameters are the thicknesses of the layers and the time-dependent dielectric permittivity of the PMMA-SP(MC) layer. The latter is modelled as $\epsilon_{donor}(\omega, t_{exp}) = \gamma(t_{exp})\epsilon_{MC}(\omega) + [1 - \gamma(t_{exp})]\epsilon_{SP}(\omega)$, where $\epsilon_{MC}$, $\epsilon_{SP}$ and $\epsilon_{donor}$ are the dielectric permittivity functions for PMMA films containing MC, SP, and their mixture, while $\gamma$ is related to the fraction of MC molecules in the SP/MC mixture. The angle-resolved transmission spectra are also analyzed in order to obtain Hopfield coefficients. Using a coupled-oscillators model (see SI), we fit solutions of the model to experimentally observed transmission peaks, which allow for extraction of the coupling between the cavity and the excitonic transitions of the emitters.

**Data availability.** The datasets generated during the current study are available from the corresponding author on reasonable request.

### Acknowledgments

Shadi Bashiri is acknowledged for AFM measurements. A. Camposeo and L. Persano acknowledge funding from the Italian Minister of University and Research through the PRIN 201795SBA3 and 20173L7W8K projects. M. Lednev, F. J. Garcia-Vidal and J. Feist acknowledge support by the European Research Council through Grant No. ERC- 2016-STG-714870 and by the Spanish Ministry for Science, Innovation, and Universities-Agencia Estatal de Investigación (AEI) through Grants RTI2018-099737-B-I00, PID2021-125894NB-I00, and CEX2018-000805-M (through the María de Maeztu program for Units of Excellence in Research and Development).

# Supplementary Information

# Active Control of Polariton-Enabled Long-Range Energy Transfer

A. Cargioli[1,2,*,#], M. Lednev[3,#], L. Lavista[1,2,#], A. Camposeo[2], A. Sassella[4], D. Pisignano[2,5], A. Tredicucci[2,5], F. J. Garcia-Vidal[3], J. Feist[3], and L. Persano[2,†]

[1]Dipartimento di Fisica "E. Fermi", Università di Pisa, Largo B. Pontecorvo 3, I-56127 Pisa, Italy

[2]NEST, Istituto Nanoscienze-CNR and Scuola Normale Superiore, I-56127 Pisa, Italy

[3]Departamento de Física Teórica de la Materia Condensada and Condensed Matter Physics Center (IFIMAC), Universidad Autónoma de Madrid, E-28049 Madrid, Spain

[4]Dipartimento di Scienza dei Materiali, Università degli Studi di Milano-Bicocca, Via Roberto Cozzi 55, I-20125 Milano, Italy

[5]Dipartimento di Fisica "E. Fermi" and Center for Instrument Sharing (CISUP), Università di Pisa, Largo B. Pontecorvo 3, I-56127 Pisa, Italy

[*]Contact Email (present address): acargioli@phys.ethz.ch

[†]Contact Email: luana.persano@nano.cnr.it

[#]These authors contributed equally to this work

# 1 Absorption and emission properties of donor and acceptor molecules

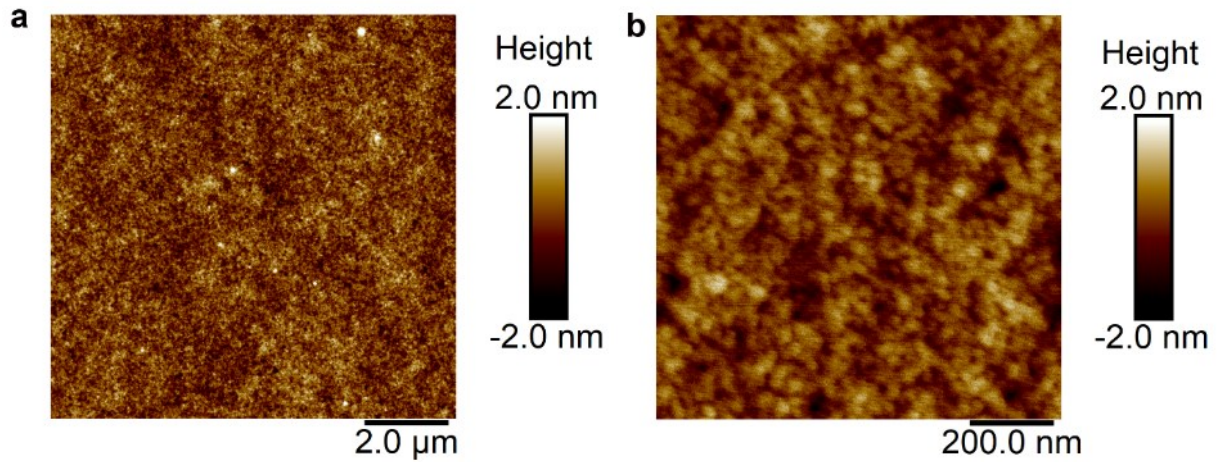

**Fig. S1: a,b.** Atomic force microscopy images, at two magnifications, of a PMMA-SP control film deposited on a PVA buffer layer. PVA is in turn deposited on a silver-coated silicon substrate. The average roughness (root mean square) is about 0.3 nm, as calculated by averaging 10 different samples area in high-magnification images (1 μm lateral size, 512 pixels/line resolution).

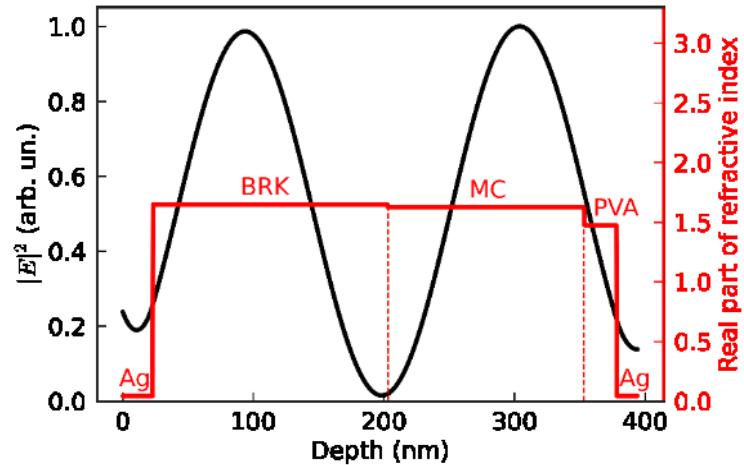

**Fig. S2:** Electric field distribution ($|E|^2$, black continuous line) and real part of the refractive index (red continuous line) along the cavity sample depth. PVA-BRK and PMMA-MC layers are here indicated as BRK and MC, respectively.

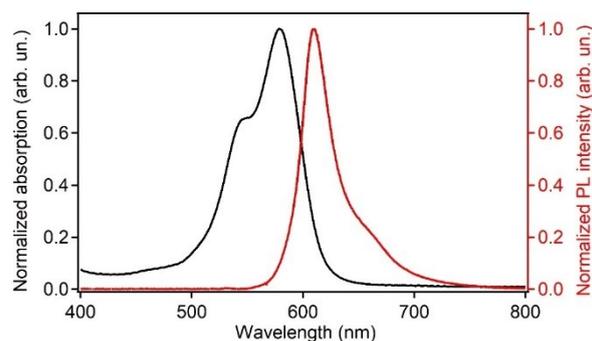

**Fig. S3: Absorption and photoluminescence of BRK.** Absorption (black continuous line, left vertical scale) and photoluminescence (PL, red continuous line and right vertical scale) spectra of a dilute solution of BRK and PVA in deionized water and methanol mixture (1:1 volumetric ratio). The weight ratio of BRK with respect to PVA is 1:1000, i.e. two orders of magnitude lower than the one used for the thin films with BRK J-aggregates. In such conditions we expect the properties of the BRK molecule to be dominant. The excitation wavelength for the emission measurement is 532 nm. The main peak of the emission spectrum of the dilute solution of BRK in PVA is at about 610 nm. The formation of J-aggregated in thin films obtained by more concentrated solutions red-shifts the absorption spectrum by about 75 nm (see also Fig.1c of the main text).

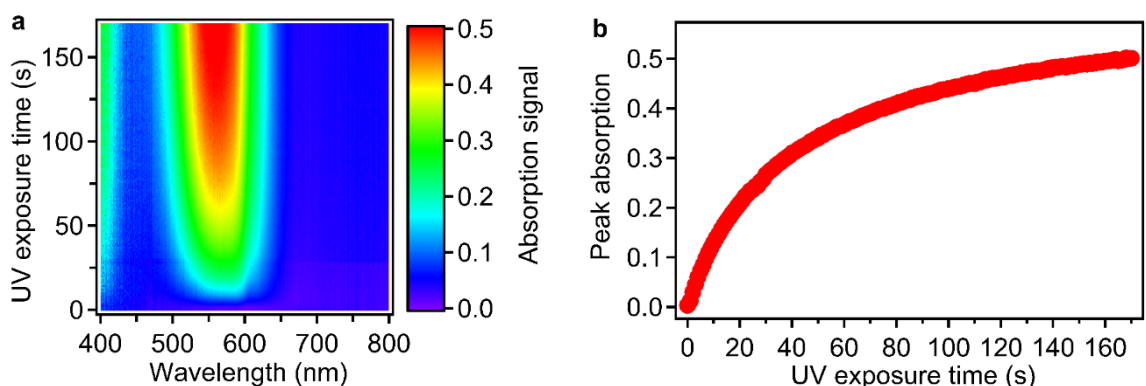

**Fig. S4: MC absorption. a** Measured absorption spectra of a MC layer upon different UV exposure times. **b** Trend of the absorption peak of the MC at 554 nm (extracted from **a**), *vs*. UV exposure time (365 nm).

## 2 Sample Fabrication

Three solutions are prepared for spinning the different layers. The first one is made by embedding the spiropyran (SP) molecules in a PMMA matrix. It contains:

- 54 mg of PMMA (Mw: 120k, Sigma-Aldrich)

- 54 mg of SP (TCI)

- 3 mL of toluene (reag. ph. pure $\geq$ 99.7%, Sigma-Aldrich).

The solution is kept in a sonicator for two hours to dissolve completely the organic materials. The second solution is made by embedding the BRK molecules in a PVA matrix. Materials used are:

- 80 mg of PVA (Mw: 124k-186k, 87-89% hydrolyzed, Sigma-Aldrich)

- 8 mg of BRK (ABCR)

- 2 mL of deionized water

- 2 mL of methanol (Carlo Erba Reagents)

After mechanical mixing, the solution is left for 3 hours on a magnetic stirrer at a temperature of 50°C. The last solution consists of bare PVA, and it is prepared according to the following procedure. Materials used are:

- 50 mg of PVA (Mw: 85k-124k, 99% hydrolyzed, Sigma-Aldrich)

- 4.95 mL of deionized water

After mechanical mixing, the solution is left for 3 hours on the magnetic stirrer at a temperature of 160 °C.

The complete multilayer is obtained by spin coating three films for 60 s on top of each other and on top of a thermally-evaporated Silver mirror (25 nm-thick). Solutions are cast in the following order: PVA solution at 4000 rpm, PMMA-SP solution at 1500 rpm, PVA-BRK solution at 2400 rpm. The cavity is completed by evaporating a top Silver mirror. An example of a sample at three different stages of the process is reported in Fig. S5a-c. A transmission measurement of a Silver mirror evaporated on quartz is shown in Fig. S5d.

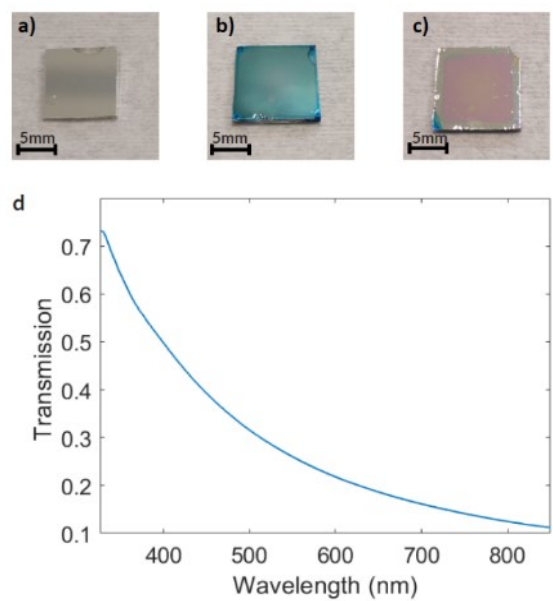

**Fig. S5: Microcavity fabrication.** Photograph of samples at different stages of fabrication. **a** Silver mirror evaporated on a quartz substrate. **b** Active region embedding both donor and acceptor molecules spin-coated on Silver. **c** Full cavity after the evaporation of 25 nm-thick Silver on top of the active region. **d** Measured optical transmission spectrum of a 25 nm-thick Silver mirror deposited on quartz.

## 3 Transmission of the microcavity at intermediate times

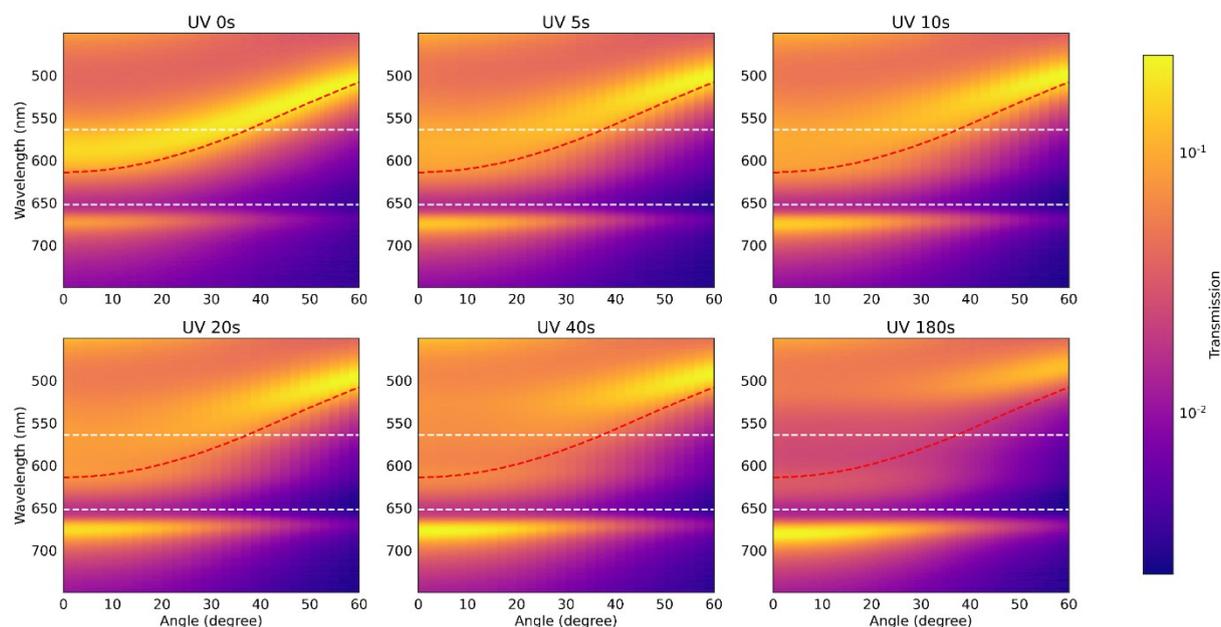

**Fig. S6: Cavity transmission measurements.** Angle-resolved transmission measurement as a function of UV light exposure time. In the chosen colorscale, unity stands for total transmission. In each colormap the bare cavity mode (red dashed line), the MC excitonic transition (upper white dashed line) and the BRK excitonic transition (lower white dashed line) are also reported.

# 4 Characterization of BRK and MC first-order cavities

The complex refractive index of the acceptor layer is derived from the characterization of a first-order cavity with PVA-BRK (thickness of the bottom and top Ag mirrors about 25 nm, thickness of the PVA-BRK about 165 nm). The measured and simulated transmission maps are reported in Fig. S7a-b. Using a Transfer Matrix Method (TMM) approach we minimize the difference between the experimental and calculated angle-resolved transmission and reflection spectra. For this procedure we vary as free parameters the layers thicknesses and the parameters for the dielectric permittivity of the active layer. In particular, we approximate the imaginary part of the dielectric permittivity of PVA-BRK as a superposition of 2 Voigt profiles (corresponding to J-aggregates and non-aggregated monomers, respectively). The Voigt profile is a convolution of Gaussian and Lorenzian profiles representing inhomogeneous and homogeneous broadening, respectively. For the dielectric permittivity of BRK we hence use 8 free parameters (2 central frequencies, 2 Gaussian widths, 2 Lorentzian widths, 2 amplitudes). The real part of the permittivity is then calculated by use of Kramers-Kronig relations. As a constant background permittivity, data available in Ref. [1] for the PVA host matrix are used. The obtained refractive index of the PVA-BRK is reported in Fig. S7c-d.

The calculated values are also compared with those obtained from the fit of ellipsometry measurements performed on a PVA-BRK layer spin-coated on a silicon/silicon oxide substrate. The contribution of several oscillators is considered, as indeed visible in the low-wavelength tail of both the $n$ and $k$ spectra. We find a good agreement between the values of the refractive index obtained by the two different methodologies.

J-aggregates are complex structures and many of their properties strongly depend on the environmental conditions, including the fabrication procedure. The simulations of more complex experiments involving donor and acceptor layers are hence performed by using the refractive index values of PVA-BRK calculated by the TMM fitting, which comes from materials undergone the same cavity fabrication processes.

A characterization of a first-order cavity with only PMMA-SP/MC is also performed. The active region consists of two 25 nm thick PVA buffer layers (in contact with the 25 nm thick Ag mirrors) and a central layer of PMMA-SP/MC matrix with a thickness about 110 nm. The experimental and simulated transmission maps are reported in Fig. S8a-b. Analogously to the procedure described above for BRK, we approximate the dielectric permittivity of PMMA-MC as a superposition of two Voigt profiles, with a constant background permittivity due to the PMMA host matrix taken from Ref. [2]. Then, we fit angle-resolved spectra using the TMM and, through a best fit procedure, we retrieve the optimal parameters that allow us to get the refractive index of the PMMA-MC (Fig. S8c-d). The permittivity of PMMA-SP in the relevant spectral range is simulated as the permittivity of PMMA with a small constant absorption. The latter value is obtained through fitting of the corresponding transmission spectra of the cavity with SP layer. The curve obtained from the fit of the refractive index of the PMMA-SP and PMMA-MC is further exploited for simulations of more complex experiments involving donor and acceptor layers, as described in Methods of the main manuscript.

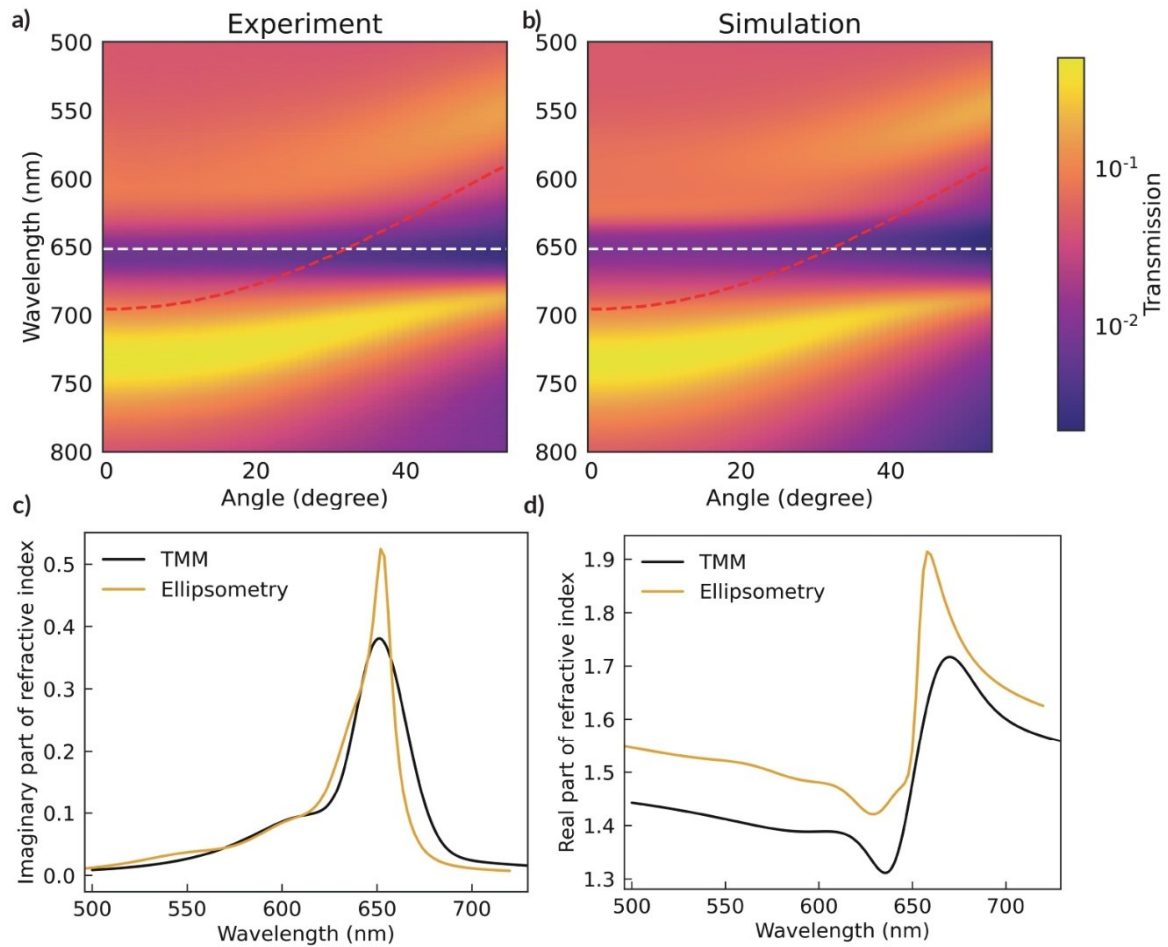

**Fig. S7: BRK cavity transmission and PVA-BRK refractive index.** Experimental **a** and simulated **b** angle-resolved transmission maps of a first-order cavity embedding a PVA-BRK layer. Imaginary **c** and real **d** part of the BRK layer refractive index, as calculated by minimizing the difference between the experimental and calculated angle-resolved transmission and reflection spectra of the first-order cavity (TMM, black line) and by ellipsometry data (yellow line).

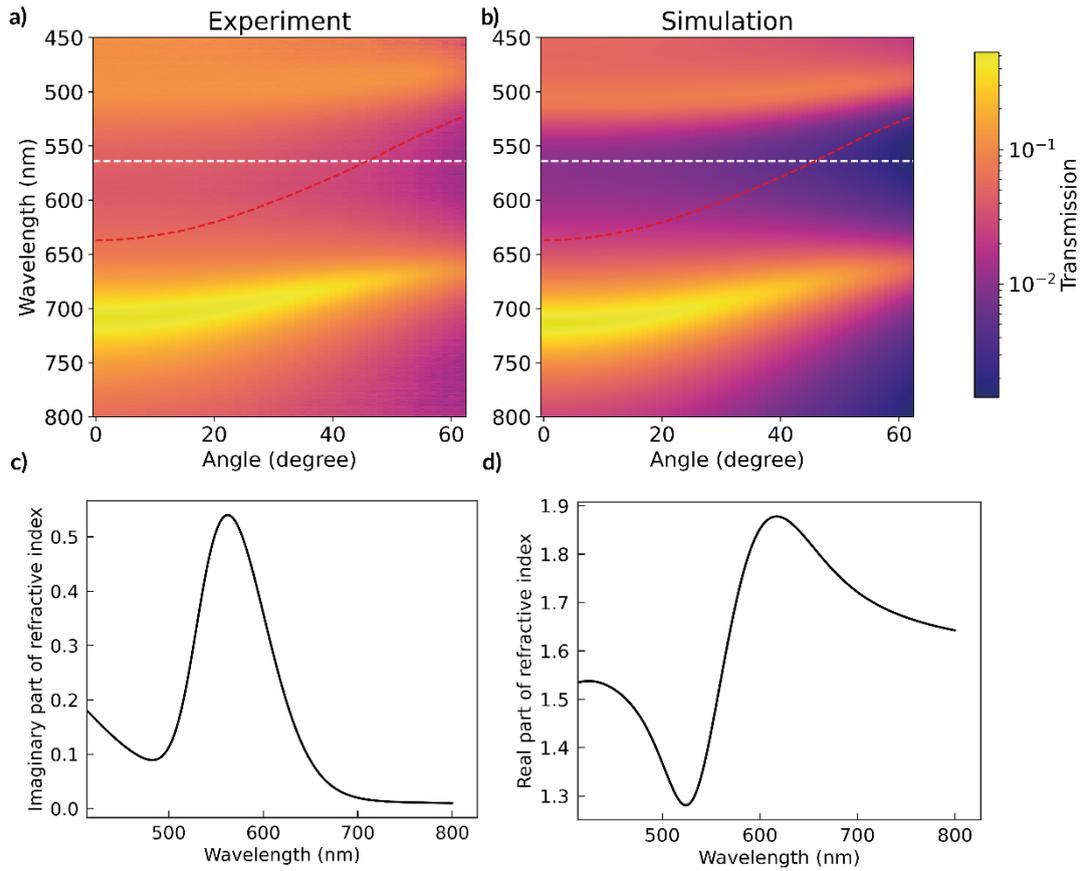

**Fig. S8: MC cavity transmission and PMMA-MC refractive index.** Experimental **a** and simulated **b** angle-resolved transmission maps of a first-order cavity embedding only the PMMA-MC layer. Fitted imaginary **c** and real **d** part of the PMMA-MC layer refractive index.

## 5 Coupled Oscillators Model

Each polaritonic dispersion is fitted with a coupled oscillators model for each UV exposure time. The method consists in solution of the eigenvalue problem:

$$
\begin{bmatrix}
\omega_{\text{BRK}} & 0 & g_{\text{BRK}} \\
0 & \omega_{\text{MC}} & g_{\text{MC}}(t_{\text{exp}}) \\
g_{\text{BRK}} & g_{\text{MC}}(t_{\text{exp}}) & \omega_{\text{cav}}(\theta)
\end{bmatrix}
\begin{bmatrix}
\alpha_{\text{BRK}}(\theta, t_{\text{exp}}) \\
\alpha_{\text{MC}}(\theta, t_{\text{exp}}) \\
\alpha_{\text{cavity}}(\theta, t_{\text{exp}})
\end{bmatrix}
= \omega_{\text{pol}}(\theta, t_{\text{exp}})
\begin{bmatrix}
\alpha_{\text{BRK}}(\theta, t_{\text{exp}}) \\
\alpha_{\text{MC}}(\theta, t_{\text{exp}}) \\
\alpha_{\text{cavity}}(\theta, t_{\text{exp}})
\end{bmatrix}
\tag{S1}
$$

where $\omega_{BRK}$, $\omega_{MC}$ and $\omega_{cav}$ are frequencies of the BRK transition, the MC transition and the cavity mode, respectively; $g_{BRK}$ and $g_{MC}(t_{\text{exp}})$ are coupling strengths between the emitters and the cavity mode; $\theta$ is the measurement angle. The eigenfrequencies $\omega_{pol}(\theta, t_{exp})$ represent the frequencies of hybrid polaritonic states, and eigenvectors carry information about Hopfield coefficients which are equal to: $|\alpha_i|^2$ ($i$=BRK, MC, cavity). The peak positions of the emitter absorption spectra are taken as their transition frequencies. The frequency of the bare cavity mode is retrieved from TMM simulations. The estimated Hopfield coefficients for each polaritonic branch for three different exposure times are shown in Figures S9, S10, S11.

In order to get coupling strengths corresponding to the experiment we perform one fitting procedure for all the exposure times $t_{exp}$, where we vary $g_{BRK}$ and $g_{MC}$ trying to minimize the difference between $\omega_{pol}(\theta, t_{exp})$ and the spectral position of the polaritonic states taken from experimental angle-resolved transmission spectra. It should be noted that for every exposure time we take different coupling strength for the donor (since we expected an increase of the number of MC molecules) and kept fixed the coupling strength between acceptor molecules and cavity mode. The dependence of $g_{MC}(t_{exp})$ is reported in the Fig. S12.

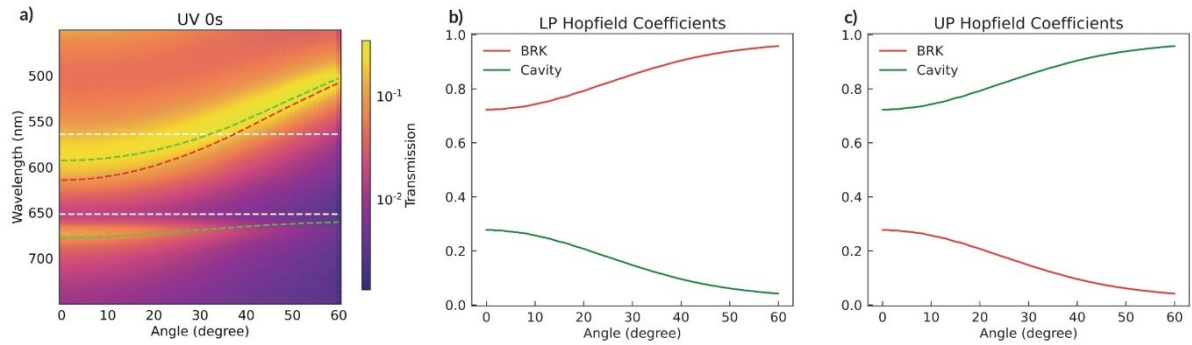

**Fig. S9: Pristine cavity transmission and Hopfield coefficients. a** Colormap of the microcavity transmission at 0 s of UV exposure and relative angular dispersion of the Hopfield coefficients for the LP **b** and UP **c** branches. The colorscale is chosen to be logarithmic. In the colormap the bare cavity mode (red dashed line), the MC excitonic transition (upper white dashed line) and the BRK excitonic transition (lower white dashed line) are also reported. The green dashed lines are the result of a fit using the coupled oscillators model.

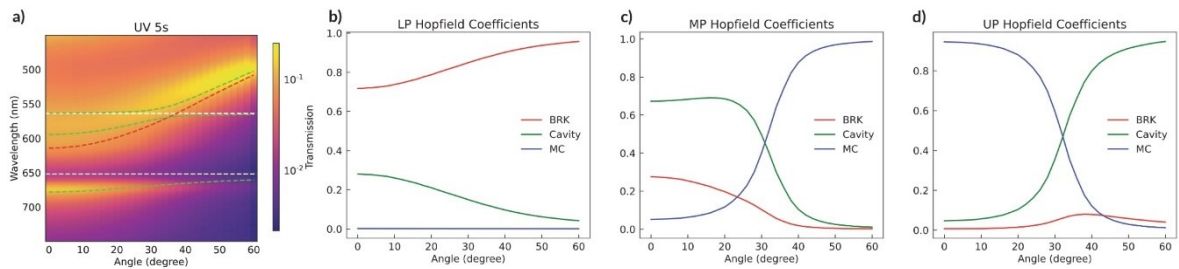

**Fig. S10: Cavity transmission at 5 s and Hopfield coefficients. a** Colormap of the microcavity transmission at 5 s of UV exposure and relative angular dispersion of the Hopfield coefficients for the LP **b**, MP **c** and UP **d** branches. In the colormap the bare cavity mode (red dashed line), the MC excitonic transition (upper white dashed line) and the BRK excitonic transition (lower white dashed line) are also reported. The green dashed lines are the result of a fit using the coupled oscillators model.

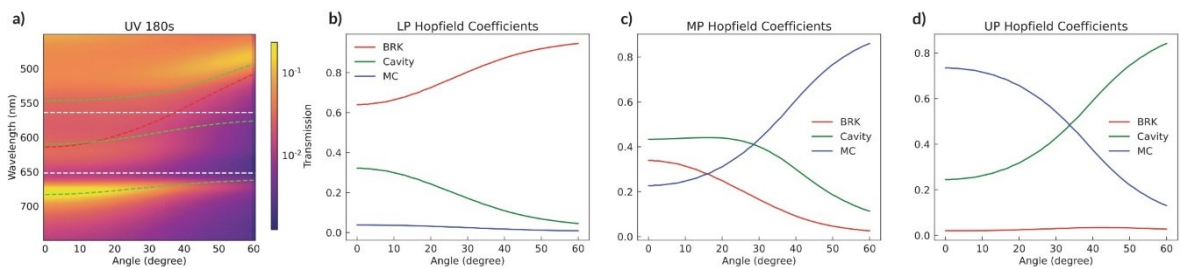

**Fig. S11: Cavity transmission at 180 s and Hopfield coefficients. a** Colormap of the microcavity transmission at 180 s of UV exposure and relative angular dispersion of the Hopfield coefficients for the LP **b**, MP **c** and UP **d** branches. In the colormap the bare cavity mode (red dashed line), the MC excitonic transition (upper white dashed line) and the BRK excitonic transition (lower white dashed line) are also reported. The green dashed lines are the result of a fit using the coupled oscillators model.

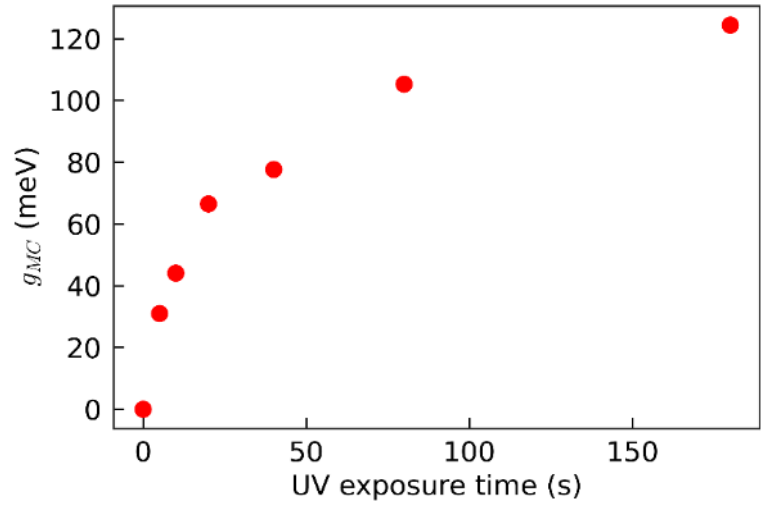

**Fig. S12: MC coupling coefficient.** Collective light-matter coupling strength of the MC molecules as a function of the UV exposure time.

## 6 Analysis of the MC to SP back-conversion

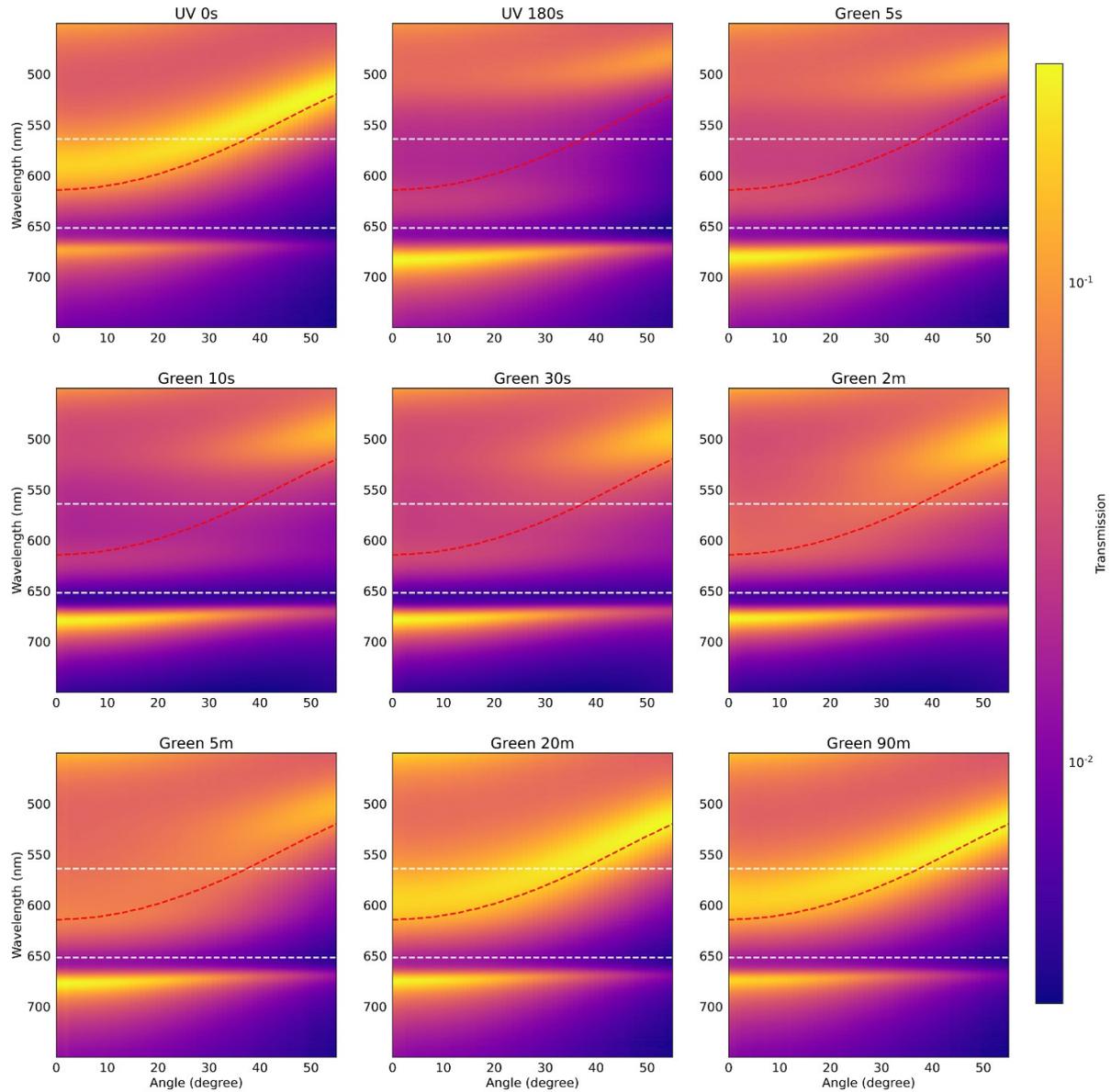

**Fig. S13: Cavity transmission during MC back-switching.** Angle-resolved transmission spectra as a function of UV and green light exposure time. In each colormap the bare cavity mode (red dashed line), the MC excitonic transition (upper white dashed line) and the BRK excitonic transition (lower white dashed line) are also reported.

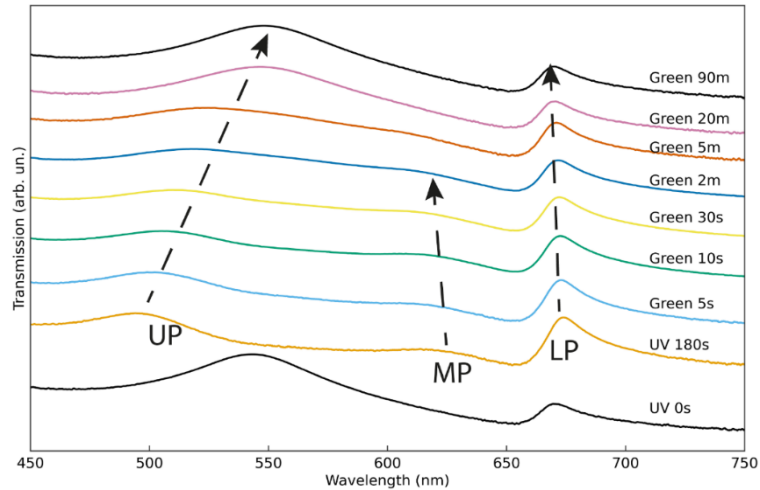

**Fig. S14: Cavity transmission spectra at fixed angle.** Transmission spectra measured at 39° incidence angle as a function of UV and green light exposure times. The black arrows highlight the shift of the UP, MP and LP branches, respectively. The spectra are vertically shifted for better clarity.

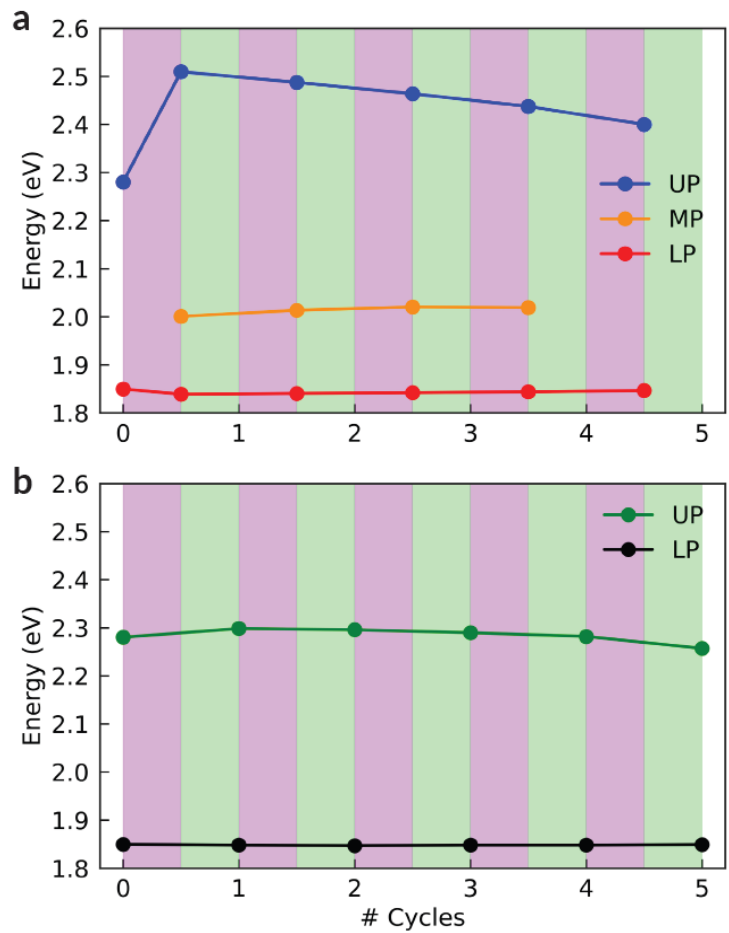

**Fig. S15: Polariton behaviour upon UV/green irradiation cycles.** Energy of the UP, MP and LP branches after repeated UV/green irradiation. The energy values are obtained from transmission spectra collected at 39° after each UV (**a**) or green (**b**) exposure step.

# 7 PL of the microcavity at intermediate UV exposure times

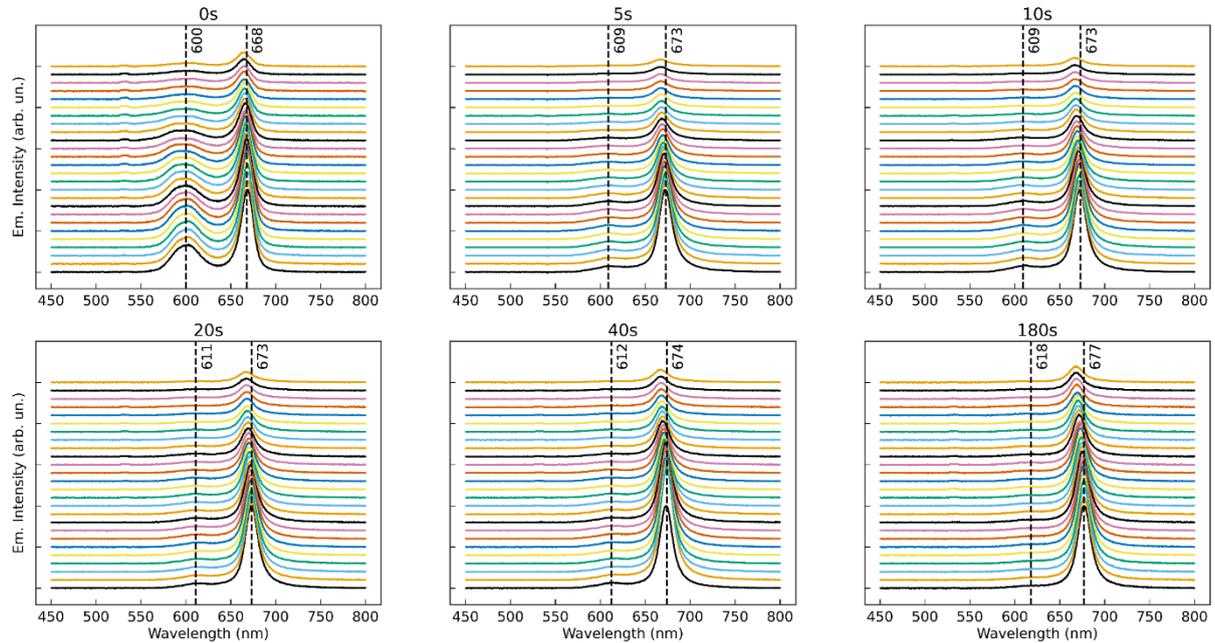

**Fig. S16: Cavity emission during SP to MC conversion.** Experimental angle-resolved emission spectra of the cavity as a function of the initial UV light exposure time. The excitation wavelength for the emission measurement is 532 nm. Each spectrum is measured every 2° starting from the bottom of the plot. For each UV exposure time, spectra are normalised to the maximum emission intensity of the one measured at 0°. The vertical dashed lines identify the peaks wavelength (nm) at 0° of emission.

## 8 Analysis of the cavity PL upon MC-to-SP back-switching

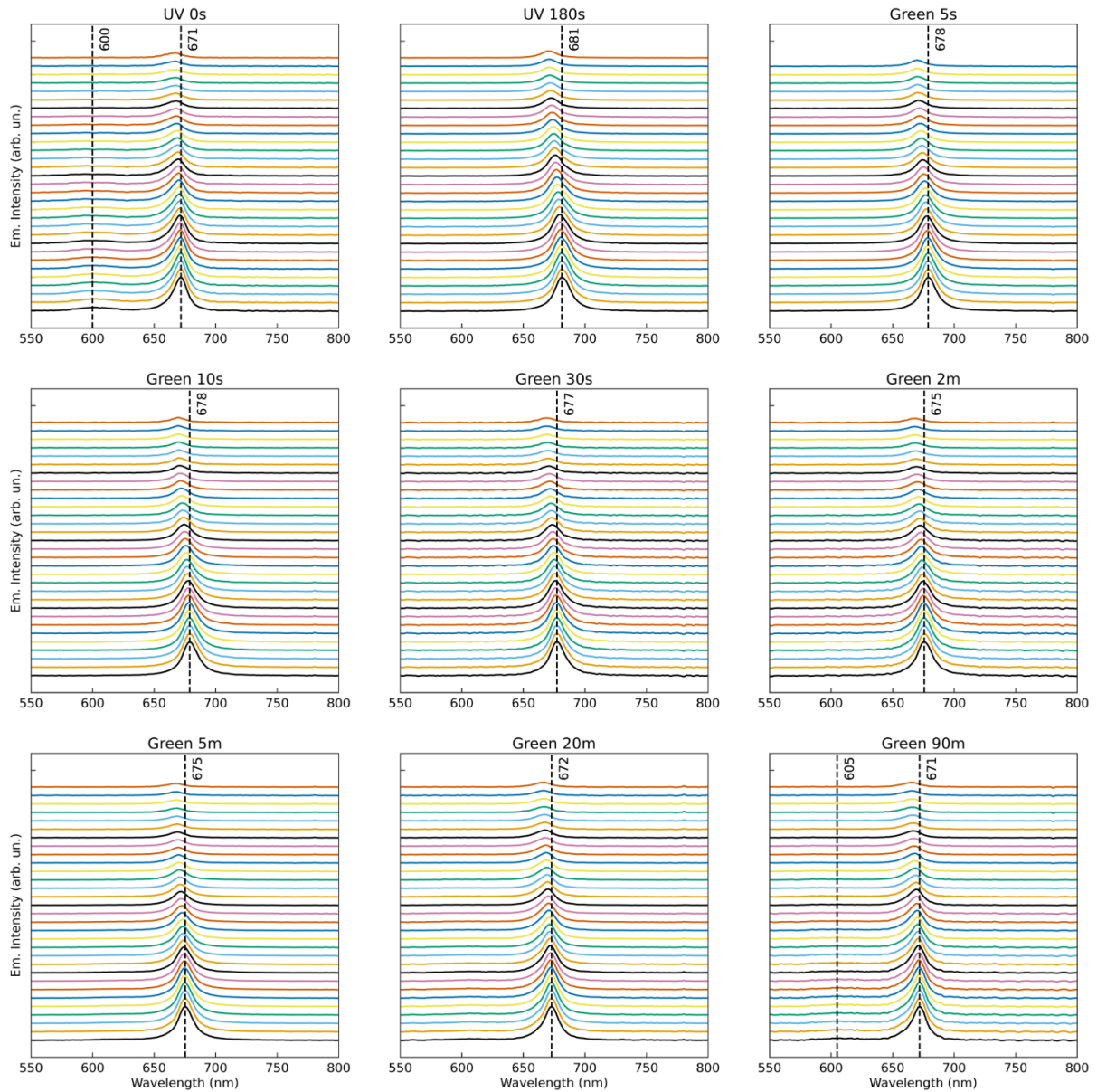

**Fig. S17: Cavity emission during MC back-switching.** Experimental angle-resolved PL spectra of the cavity as a function of the initial UV and subsequent green light exposure times. Excitation wavelength: 532 nm. Spectra are vertically shifted. Each spectrum is measured at a given angle value, each with 2° spacing from the previous one, starting from the bottom of the plot. The spectra are all normalised to the maximum emission intensity of the 0° spectrum. The dashed vertical lines identify the peaks wavelength at 0°.

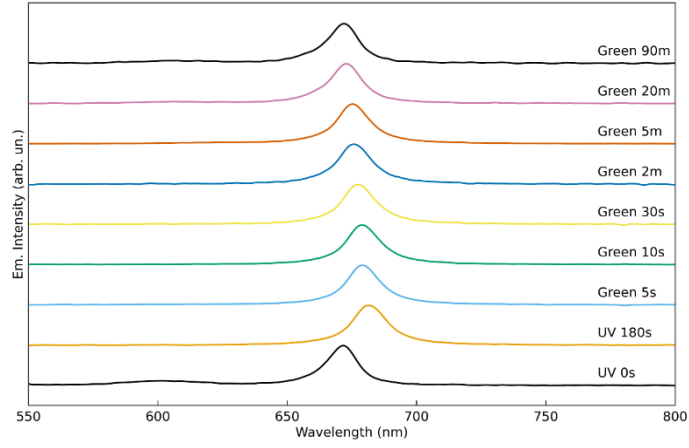

**Fig. S18: Cavity emission during MC back-switching at fixed angle.** PL spectra of the cavity as a function of the initial UV and subsequent green light exposure times measured at 0°. Excitation wavelength for the emission measurements: 532 nm. Each spectrum is normalized to its maximum intensity value. The spectra are vertically shifted for better clarity.

## 9 The model for the cavity emission

We assume that the cavity emission signal can be regarded as emission of each molecular species modulated by an effective filtering induced by the cavity. Thus, we use the following expression:

$$I_{cav}(\omega, \theta; t_{exp}) = \alpha(t_{exp}) F_{BRK}(\omega, \theta; t_{exp}) I_{BRK}(\omega; t_{exp}) +$$

$$\beta(t_{exp}) F_{SP/MC}(\omega, \theta; t_{exp}) I_{SP/MC}(\omega; t_{exp}) \tag{S2}$$

where $F_i$($i$=BRK, SP/MC) is the filter function for the emission of molecular layer $i$. Since the emission of each active layer goes from the middle of the sample to the detector located behind one of the mirrors, for a correct estimation of $F_i$, the corresponding transmission coefficient should be considered. So $F_i$ can be expressed as:

$$F_i(\omega, \theta; t_{exp}) \propto \int_{\substack{i-th \\ layer}} T_{x \to det}(x; \omega, \theta, t_{exp}) \, dx \tag{S3}$$

where $T_{x \to det}$ represents the transmission of the light from the plane $x$ of the layer $i$ to the detector. This quantity can be easily calculated by TMM. Since $T_{x \to det}$ strongly depends on the position of the source $x$, we (incoherently) integrate the emission contributions coming from the different parts of the layer $i$. The results obtained with the expression for $F_i$ are presented in Fig. S19.

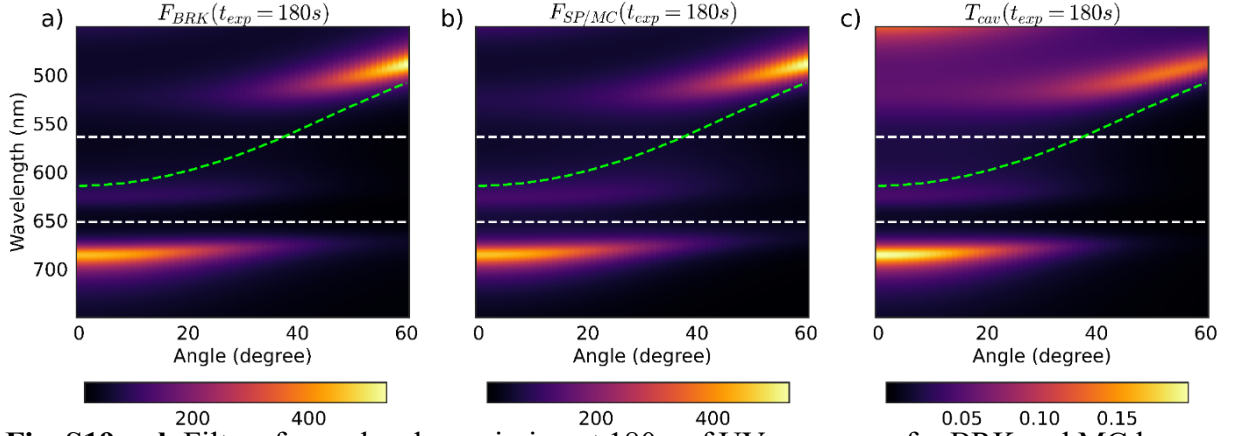

**Fig. S19: a,b** Filters for molecular emission at 180 s of UV exposure, for BRK and MC layers, respectively. **c** Simulated transmission of the cavity at $t_{exp} = 180$ s. In each colormap the bare cavity mode (green dashed line), the MC excitonic transition (upper white dashed line) and the BRK excitonic transition (lower white dashed line) are also reported.

It can be seen that for both molecular species, the maps are similar to the total cavity transmission. The main deviations are in the short-wavelength region, which is off-resonant from the emission of the molecular species. We also note that the scale of BRK and SP/MC filter functions are similar, so that the cavity transmission is a good approximation for the emission filter for both molecular species. We also note that the difference in absolute scale between the filter functions and cavity transmission does not play any role, since for the analysis we use normalized weight coefficients.

## 10 Rate equations model

In order to validate the approach that we use to describe emission dynamics for our system, we also implement a rate equations model for the lower polariton branch (LPB)[3]. Since the photochemical reaction has a rate comparatively small to the polariton dynamics we can a use steady state condition for the population of the LPB:

$$\frac{dN_{\text{LPB}}}{dt} = 0 = C_1 \left|\alpha_A^{LPB}\right|^2 + C_2 \left|\alpha_D^{LPB}\right|^2 + C_3 P_{\text{LPB}} \left|\alpha_{\text{ph}}^{LPB}\right|^2$$
$$- C_4 N_{\text{LPB}} \left|\alpha_{\text{ph}}^{LPB}\right|^2 - C_5 N_{\text{LPB}} \left|\alpha_A^{LPB}\right|^2 - C_6 N_{\text{LPB}} \left|\alpha_D^{LPB}\right|^2 \quad \text{(S4)}$$

where $N_{LPB}$ is the lower polariton branch population, $\left|\alpha_i^{LPB}\right|^2$ ($i$=A, D, ph) is the LPB Hopfield coefficient for acceptor, donor and cavity photon respectively, and $P_{LPB}$ is the overlap between LPB and emission spectra of the bare molecules. In this equation, the first two terms describe the processes of vibrational scattering from donor and acceptor excitonic reservoirs, the third term denotes the radiative pumping, the fourth one stands for radiative decay through the photonic component of lower polariton, and the last two terms describe relaxation to excitonic reservoirs.

We assume that for our system radiative pumping dominates over vibrational scattering, since the LPB significantly overlaps with the emission spectra of both molecular species. Also, we disregard the last term in the equation, assuming that its contribution to the LPB population is negligible compared to the radiative mechanism and LP relaxation to the acceptor reservoir, in full accordance with previous studies[3-5].

$$N_{LPB} = \frac{C_1|\alpha_A^{LPB}|^2 + C_2|\alpha_D^{LPB}|^2 + C_3 P_{LPB}|\alpha_{ph}^{LPB}|^2}{C_4|\alpha_{ph}^{LPB}|^2 + C_5|\alpha_A^{LPB}|^2 + C_6|\alpha_D^{LPB}|^2} \approx \frac{C_3 P_{LPB}|\alpha_{ph}^{LPB}|^2}{C_4|\alpha_{ph}^{LPB}|^2 + C_5|\alpha_A^{LPB}|^2} \quad \text{(S5)}$$

Since the emission intensity $I_{LPB} \propto N_{LPB}|\alpha_{ph}^{LPB}|^2$, we can write:

$$I_{LPB} \propto \frac{P_{LPB}|\alpha_{ph}^{LPB}|^2}{1 + R}, \quad \text{(S6)}$$

where $R = \frac{C_5|\alpha_A^{LPB}|^2}{C_4|\alpha_{ph}^{LPB}|^2}$ is the ratio between the rates defining LP relaxation into acceptor reservoir and LP radiative decay, respectively. This ratio can be estimated using TMM. In order to obtain $R$ we performed simulations calculating the power emitted into free space and absorbed in the acceptor molecules for emission from the donor molecules. In this scenario, we can estimate the rate of LP decay as the power emitted when the sources are distributed over the donor layer ($P_{out}$). LP relaxation to the acceptor reservoir then corresponds to the absorption of light emitted from the donor layer ($A_{acc}$). The latter process is mediated by radiative modes of the hybrid system (polaritons) and, thus, corresponds to the cascaded mechanism we aim to describe (donor → LP,MP,UP → acceptor). However, since only the LP is close to resonant with the acceptor reservoir, the acceptor absorption is mostly mediated by the LP. The resulting frequency-dependent rates for the case of 0 degree, 180s of UV illumination are presented in Fig. S20. The integrated ratio $R = 1.19$ for this case, which means that in our system LP relaxation to the acceptor reservoir is of the same order as LP radiative decay.

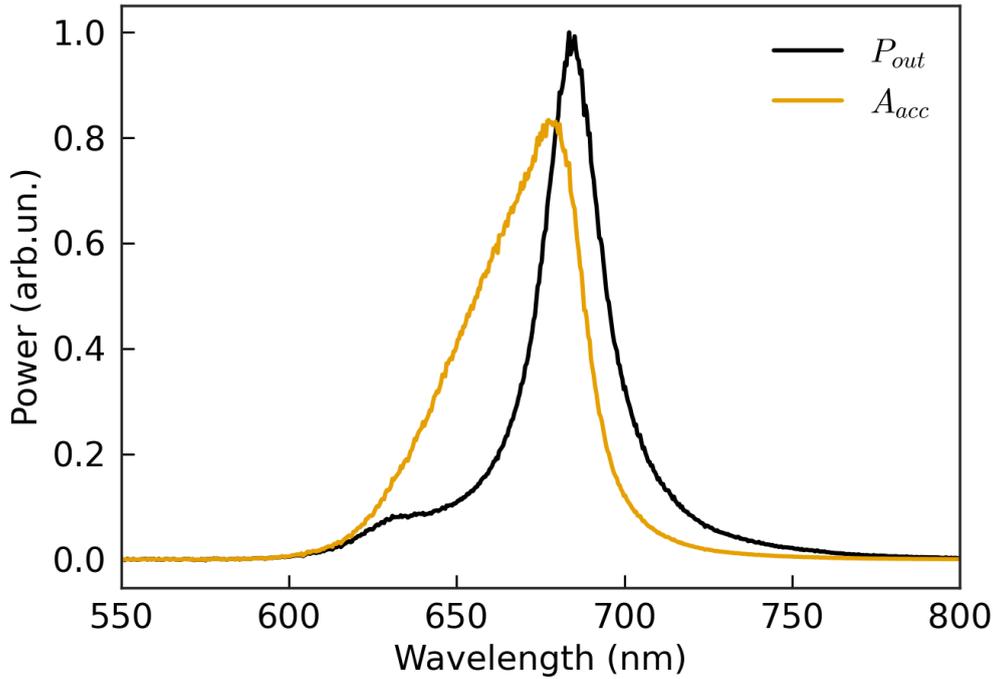

**Fig. S20:** Power leaving the cavity due to donor emission (black), acceptor absorption following the donor emission (yellow) at 0 degree, 180s of UV illumination.

The equation S6 is equivalent to the approach we use for analysis of the emission properties of the hybrid system. The comparison of the two approaches with experimental data are reported in Fig. S21. In order to calculate the radiative pumping contribution for the rate equation approach we retrieve the LPB from the cavity transmission spectra. Since for high angles this peak is barely visible, we compute the graphs only for angles smaller than 30°. It could be noticed that the approach we used for the emission dynamics analysis is able to reproduce the same behavior as experimental data and gives qualitatively the same results as the rate equations method, which, in turn, proves the validity of aforementioned assumptions.

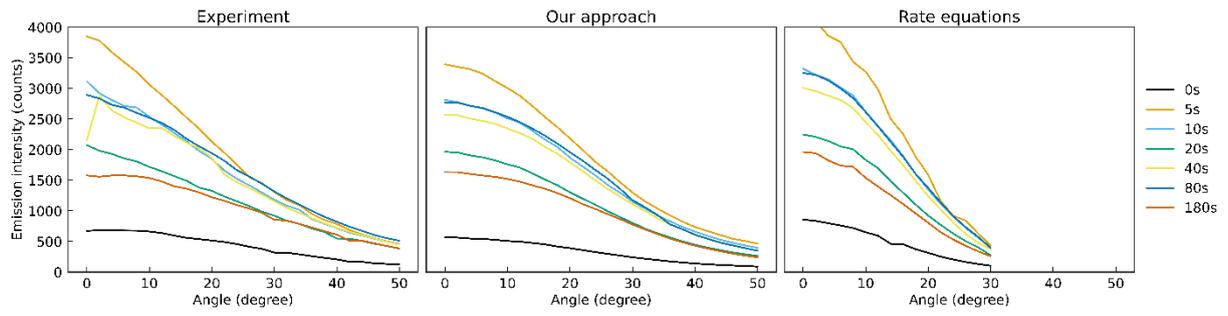

**Fig. S21: Emission intensity from lower polaritonic state for different times from the start of reaction.** Left plot: experimental data; middle plot: results obtained using our approach described in the main text; right plot: the results for rate equations approach.

## 11 PL properties of the molecules outside and inside the cavity

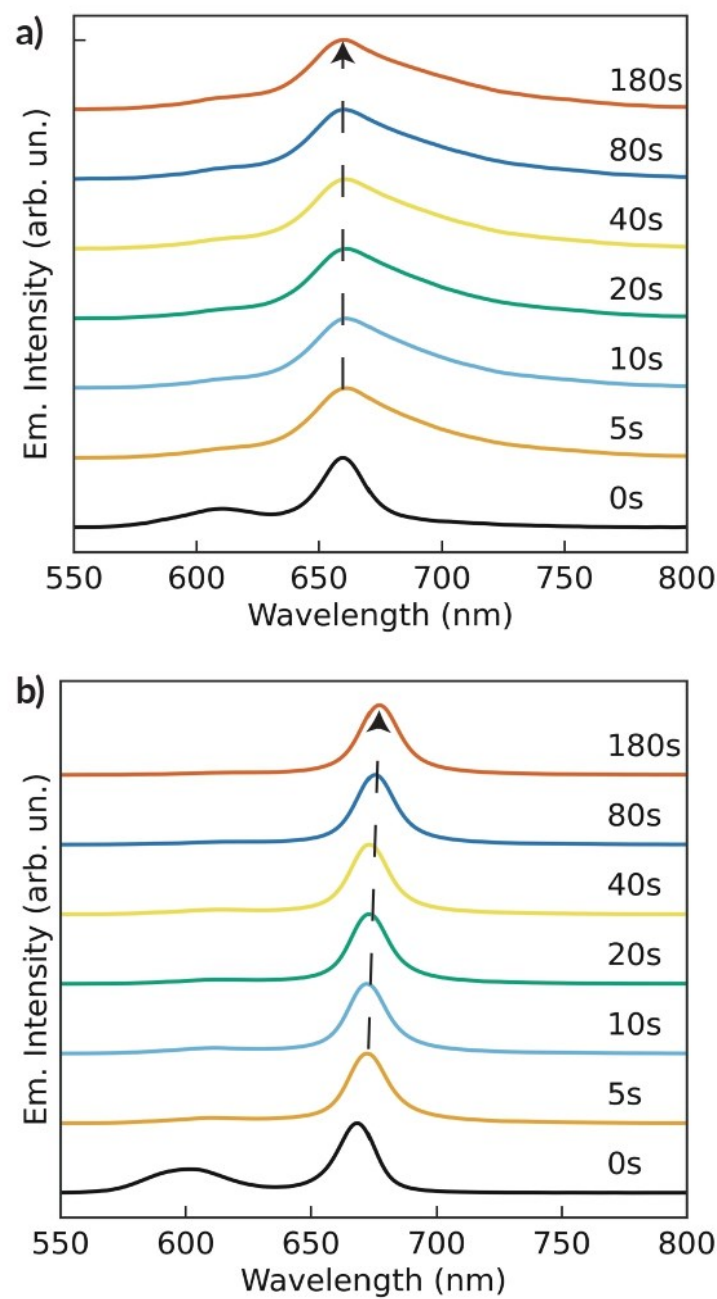

**Fig. S22:** Emission spectra of the photo-active multilayer with BRK and MC molecules out of cavity (**a**) and in the cavity (**b**), upon varying the UV exposure times between 0 s and 180 s. A green laser at 532 nm is used for PL excitation. Spectra are vertically shifted, and each spectrum is normalised to its maximum intensity value.

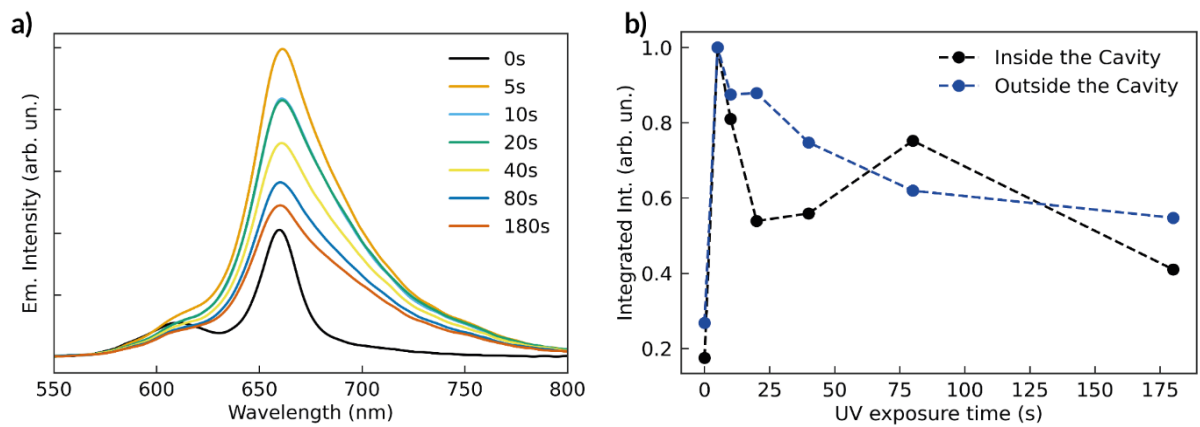

**Fig. S23: Emission properties of the photo-active multilayer with BRK and MC molecules. a** Emission spectra of the multilayer outside the cavity acquired at 0° as a function of the UV light exposure time, by using a green laser at 532 nm for PL excitation. **b** Comparison of the integrated emission intensity of the multilayer inside the cavity and outside the cavity (at 0°).

## 12 Cavity with only the acceptor

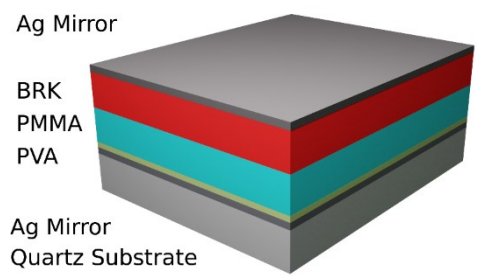

**Fig. S24:** Schematics of the realized acceptor-only cavity.

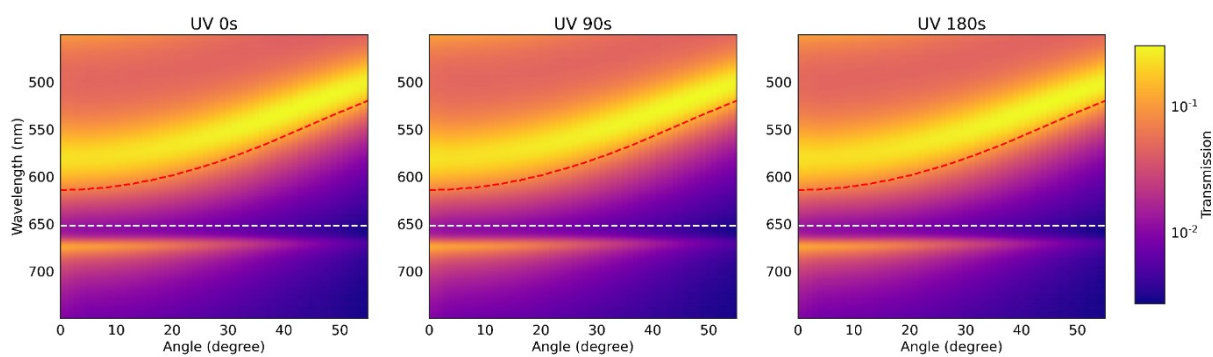

**Fig. S25: Acceptor-only cavity transmission measurements.** Angle-resolved transmission measurement as a function of UV light exposure time. In each colormap the bare cavity mode (red dashed line) and the BRK excitonic transition (white dashed line) are reported.

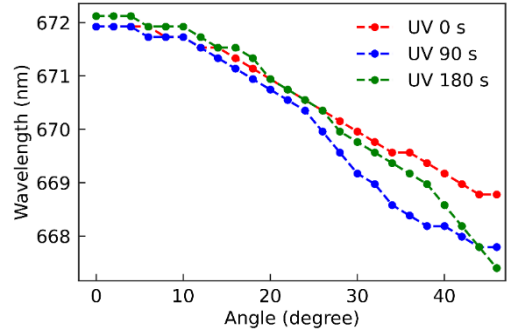

**Fig. S26: Acceptor-only cavity emission measurements.** Angular dependence of the wavelength of the cavity emission peak at different UV exposure times. The excitation wavelength for the emission measurement is 532 nm.

## 13 Cavity with only the donor

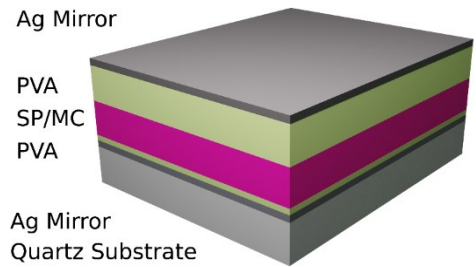

**Fig. S27:** Schematics of the realized donor-only cavity.

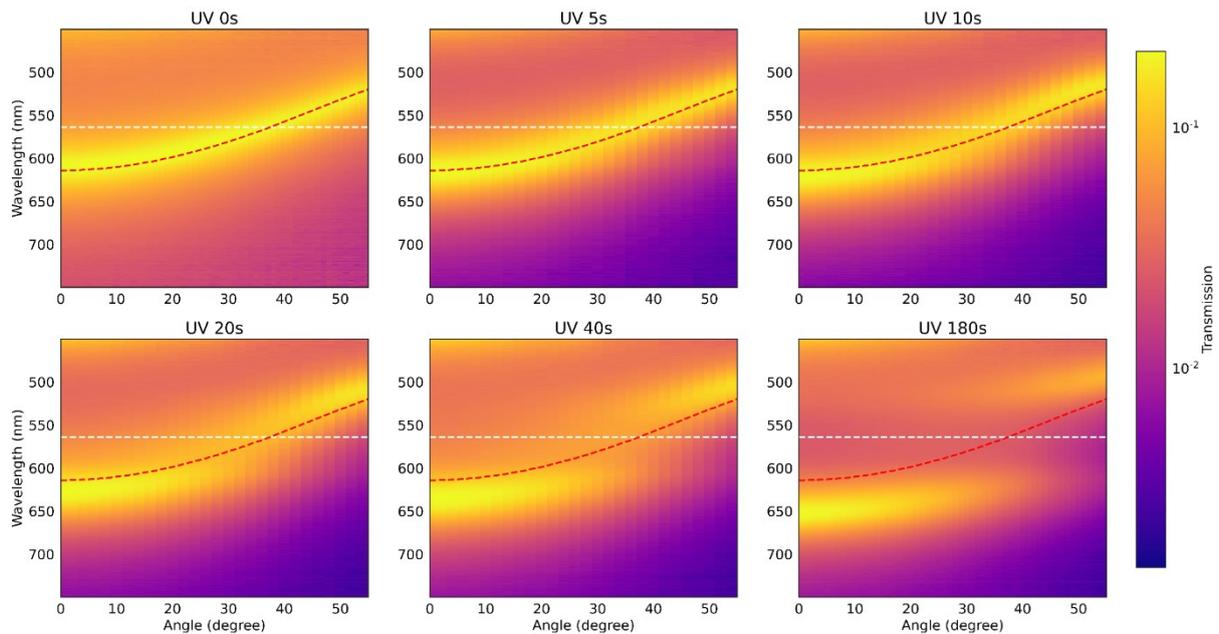

**Fig. S28: Donor-only cavity transmission measurements.** Angle-resolved transmission measurement as a function of UV light exposure time. In each colormap the bare cavity mode (red dashed line) and the MC excitonic transition (white dashed line) are reported.

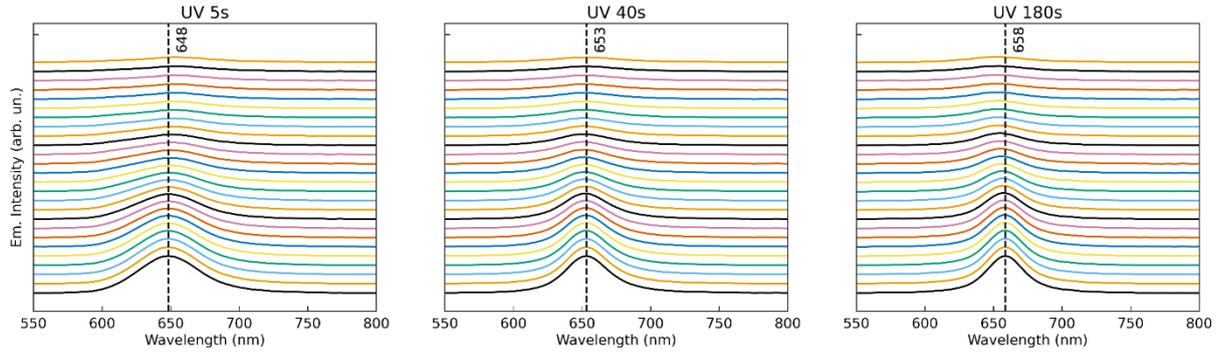

**Fig. S29: Donor-only cavity emission measurements.** Experimental angle-resolved emission spectra of the donor-only cavity as a function of the UV light exposure time. The excitation wavelength for the emission measurement is 532 nm. For each UV exposure time, spectra are normalised to the maximum emission intensity of the one measured at 0°. The vertical dashed lines identify the peaks wavelength (nm) at 0°.

## 14 Calculation of the BRK and MC contributions to the emission pattern of the multilayer outside the cavity

In order to find the donor and acceptor contributions to the emission, we approximate the PL signal of the multilayer by the following formula:

$$I_{ML}(\omega; t_{exp}) = \alpha_{ML}(t_{exp})I_{BRK}(\omega) + \beta_{ML}(t_{exp})I_{SP/MC}(\omega; t_{exp}) \tag{S7}$$

where $I_{BRK}$, $I_{SP/MC}$ are the emission intensities of the molecules outside the cavity, and $\alpha_{ML}$, $\beta_{ML}$ are phenomenological weight coefficients for BRK and MC, respectively. The coefficients and $\alpha_{ML}$, $\beta_{ML}$ are obtained by fitting the time-dependent spectra $I_{ML}(\omega; t_{exp})$ to the experimental spectra shown at the Fig. S23a using the weight coefficients $\alpha_{ML}$ and $\beta_{ML}$ as free parameters. Some examples of the results of the fitting are shown in Fig. S30.

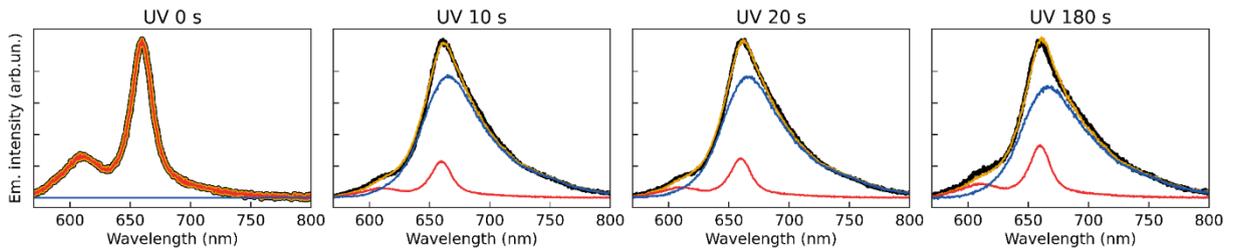

**Fig. S30: Fitting of the multilayer emission spectra outside the cavity.** Experimental (black) and simulated (yellow) emission spectra for multilayer outside the cavity for exposure time of 0 s, 10 s, 20 s, 180 s using a laser at 532 nm as a pump. BRK contribution $\alpha_{ML}(t_{exp})I_{BRK}(\omega)$ (red) and SP/MC contribution $\beta_{ML}(t_{exp})I_{SP/MC}(\omega; t_{exp})$ (blue) into the emission pattern.

As can be seen from Fig. S30, the emission spectra of the multilayer outside the cavity can be very well described as linear superposition of the spectra of separate molecular species. The temporal dynamics of the normalized weight coefficients $\alpha_n = \frac{\alpha_{ML}}{\alpha_{ML} + \beta_{ML}}$ and $\beta_n = \frac{\beta_{ML}}{\alpha_{ML} + \beta_{ML}}$ is presented in Fig. 6b of the main text.

## 15 Model sensitivity to experimental parameter variations

The intensity of the effective excitation can vary during the experiments. This variation may be related to the excitation laser or to variations in the excitation efficiency within the cavity. The latter is affected by various factors such as changes in refractive index due to UV exposure and modifications in the cavity transparency resulting from shifts in polaritonic states.

The changes of effective pumping can significantly affect the overall emission from the sample, which in turn leads to variations in the weight coefficients $\alpha$ and $\beta$ used for the emission analysis. However, both $\alpha$ and $\beta$ should be proportional to the excitation intensity $I_{exc}$, as the absorbance of molecules and, in turn, the emission intensity both depend linearly on $I_{exc}$. The subsequent normalization of the weight coefficients that we perform has the result that $\alpha_n = \frac{\alpha}{\alpha + \beta}$ and $\beta_n = \frac{\beta}{\alpha + \beta}$ do not depend on the excitation intensity. Normalizing the weight coefficients also allows for a fair comparison of results obtained inside and outside the cavity, provided that the thickness ratio of the active layers remains constant.

It is also important to note that emission of the multilayer structure, both inside and outside the cavity, is affected by the emission pattern of the SP/MC layer in PMMA. This pattern is significantly modified throughout the UV exposure due, on one side, to the increase of MC concentration and, on the other side, to the concomitant photo-bleaching effects related to photo-oxidation and other fatigue or aggregation mechanisms impacting on the merocyanine molecules[6-9], as evidenced by the emission spectra $I_{SP/MC}(\omega; t_{exp})$ (Fig. S23a). However, this does not affect the weight coefficients since these effects are already accounted for in Eq. 1 of the main text and in Eq. S7 of SI, which consider the quantity $I_{SP/MC}(\omega; t_{exp})$.

## 16 Dependence of energy transfer on cavity detuning

To investigate the underlying nature of the observed results, we performed analogous experiments with the cavity intentionally off-resonant to the molecular species. To achieve this, we realized a cavity with BRK and SP/MC layer thicknesses 240 nm and 225 nm, respectively. It is worth noting that in the resonant cavity the thicknesses are 180 nm and 150 nm. Hence the choice of layer thickness values ensures that the ratio between them remained approximately constant for a meaningful comparison. The distribution of the electric field inside the off-resonant cavity is show in Figure S31.

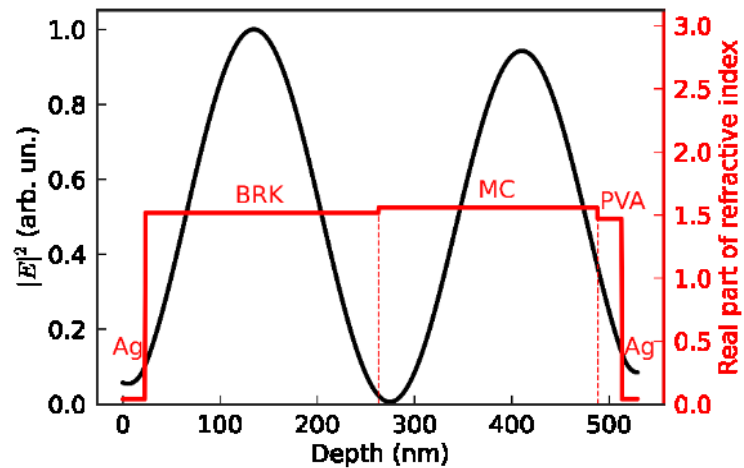

**Fig. S31:** Electric field distribution ($|E|^2$, black continuous line) and real part of the refractive index (red continuous line) along the cavity sample depth for the off-resonant cavity. PVA-BRK and PMMA-MC layers are here indicated as BRK and MC respectively.

In Figure S32, we present the transmission of the off-resonant cavity sample at 0-180 s UV exposure times. Additionally, in the Figure S33 we display the angle-resolved emission spectra of the same cavity sample.

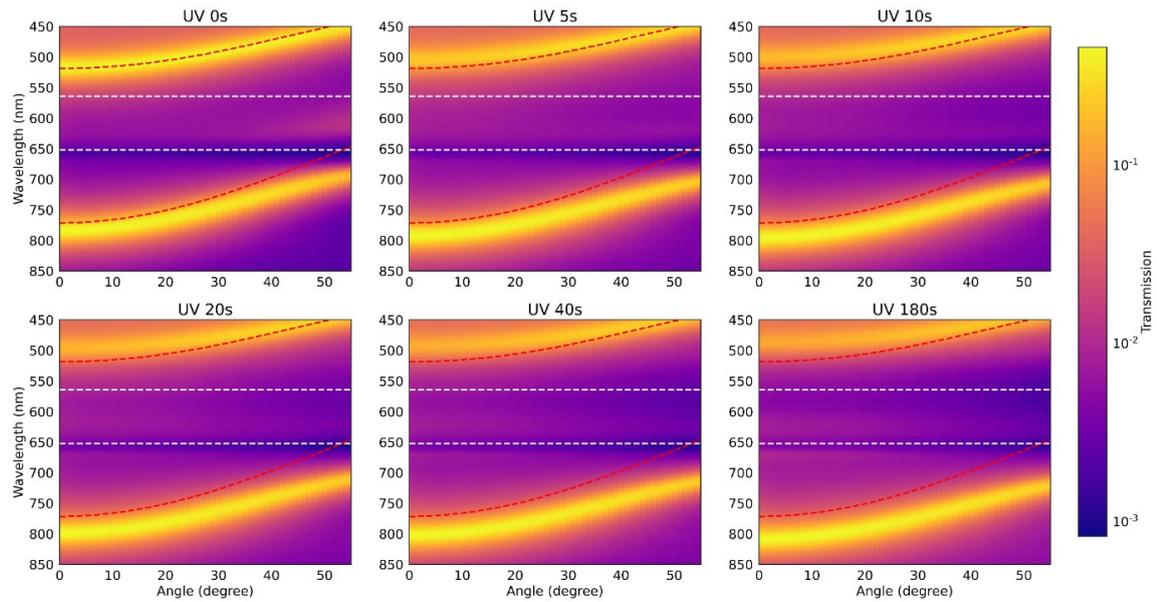

**Fig. S32**: Experimental angle-resolved transmission maps of the off-resonant cavity at different UV exposure times.

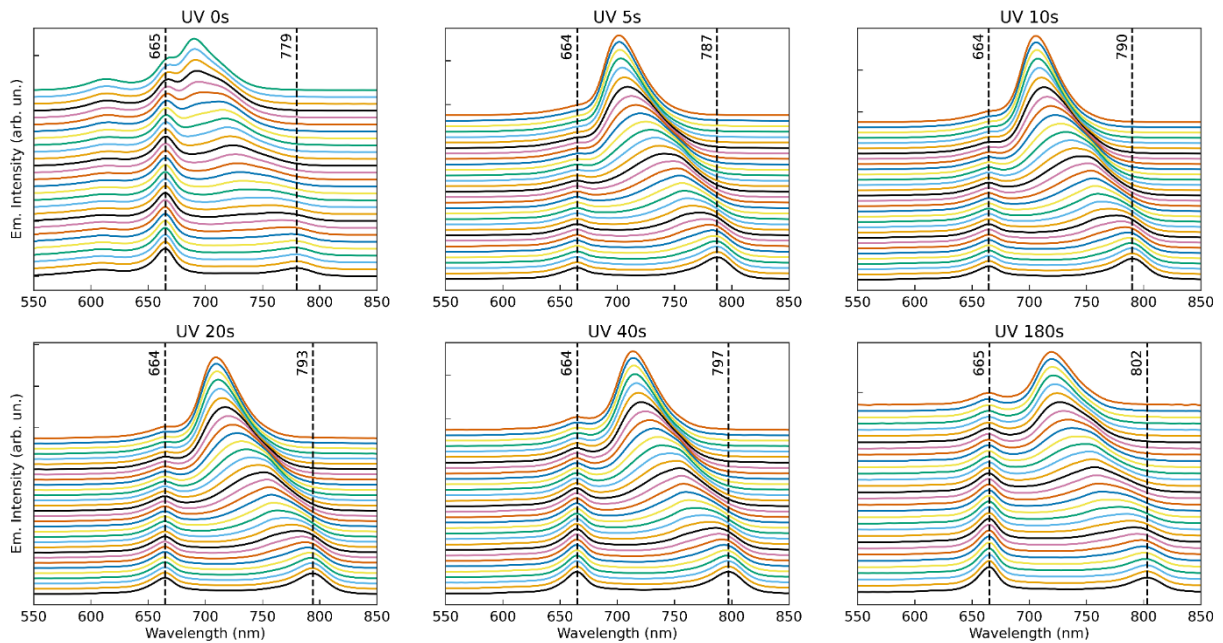

**Fig. S33**: Experimental angle-resolved emission spectra of the off-resonant cavity as a function of the UV light exposure time. Excitation wavelength: 532 nm. Each spectrum is measured at a given angle value, each with 2° spacing from the previous one, starting from the bottom of the plot. For each UV exposure time, spectra are normalised to the maximum emission intensity of the one measured at 0°. The vertical dashed lines identify the peaks wavelength (nm) at 0° of emission.

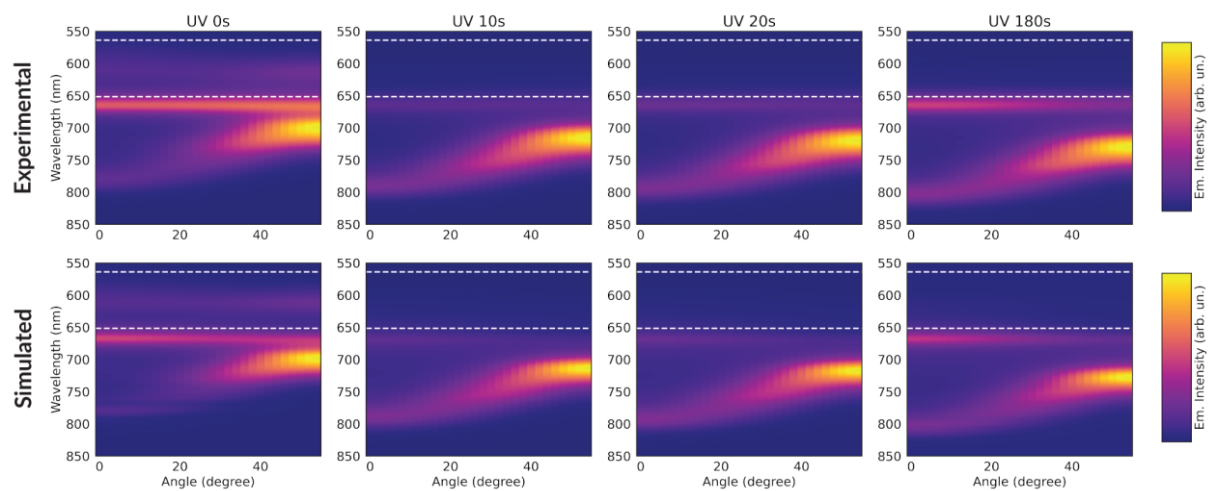

**Fig. S34: Angle-resolved PL.** Experimental and simulated angle-resolved emission maps of the cavity undergone UV exposure for 0 s, 10 s, 20 s and 180 s (from left to right) and then excited by a 532 nm pump. White dashed lines show the spectral wavelengths of donor and acceptor absorption peaks.

To fit the experimental emission data, we employed the procedure described in the main text. The results of the simulations are presented in the bottom row of Figure S34. The fitting results accurately reproduce the experimental emission maps. The corresponding weight coefficients for these simulations are shown in Figure 6e of the main manuscript. Furthermore (Figure 6d of the main manuscript), using a similar procedure, we determined the weight coefficients corresponding to BRK and SP/MC molecules for the emission from the multilayer outside the cavity (with film thicknesses corresponding to the off-resonant cavity experiment).

The weight coefficients obtained with the off-resonant cavity thicknesses exhibit a similar qualitative behavior over time for both the outside-cavity and inside-cavity cases. Moreover, for the outside-cavity scenario, the contributions of BRK and MC to the emission show similar values to those obtained in the resonant case, which can be attributed to the approximately constant ratio between the thicknesses of the active layers in both experiments.

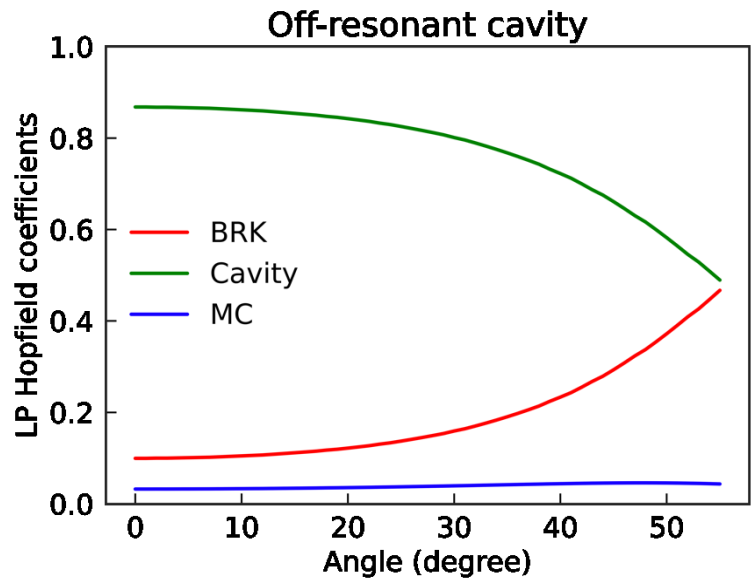

**Fig. S35:** Hopfield coefficients for the LP branch of the off-resonant cavity.